\documentclass[twocolumn]{aastex631}

\newcommand{\Msun}{\,\ensuremath{\mathrm{M}_\odot}}

\usepackage{amsmath}

\shorttitle{Machine learning detects Pop~III multiplicity}
\shortauthors{Hartwig et al.}

\begin{document}

\title{Machine learning detects multiplicity of the first stars in stellar archaeology data}

\newcommand{\SoS}{Department of Physics, School of Science, The University of Tokyo, Bunkyo, Tokyo 113-0033, Japan}
\newcommand{\ipi}{Institute for Physics of Intelligence, School of Science, The University of Tokyo, Bunkyo, Tokyo 113-0033, Japan}
\newcommand{\IPMU}{Kavli Institute for the Physics and Mathematics of the Universe (WPI), The University of Tokyo Institutes for Advanced Study, The University of Tokyo, Kashiwa, Chiba 277-8583, Japan}
\newcommand{\NAOJ}{National Astronomical Observatory of Japan, Mitaka, Tokyo 181-8588, Japan}

\correspondingauthor{Miho Ishigaki}
\email{miho.ishigaki@nao.ac.jp}

\author[0000-0001-6742-8843]{Tilman Hartwig}
\affiliation{\ipi}
\affiliation{\SoS}
\affiliation{\IPMU}

\author[0000-0003-4656-0241]{Miho N. Ishigaki}
\affiliation{\NAOJ}
\affiliation{Astronomical Institute, Tohoku University, 6-3, Aramaki, Aoba-ku, Sendai, Miyagi 980-8578, Japan}

\author[0000-0002-4343-0487]{Chiaki Kobayashi}
\affiliation{Centre for Astrophysics Research, Department of Physics, Astronomy and Mathematics, University of Hertfordshire, Hatfield AL10 9AB, UK}
\affiliation{\IPMU}

\author{Nozomu Tominaga}
\affiliation{\NAOJ}
\affiliation{Department of Physics, Faculty of Science and Engineering, Konan University, 8-9-1 Okamoto, Kobe, Hyogo 658-8501, Japan}
\affiliation{\IPMU}

\author[0000-0001-9553-0685]{Ken'ichi Nomoto}
\affiliation{\IPMU}

\begin{abstract}
In unveiling the nature of the first stars, the main astronomical clue is the elemental compositions of the second generation of stars, observed as extremely metal-poor (EMP) stars, in our Milky Way Galaxy. However, no observational constraint was available on their multiplicity, which is crucial for understanding early phases of galaxy formation. We develop a new data-driven method to classify observed EMP stars into mono- or multi-enriched stars with Support Vector Machines. We also use our own nucleosynthesis yields of core-collapse supernovae with mixing-fallback that can explain many of observed EMP stars. Our method predicts, for the first time, that $31.8\% \pm 2.3\%$ of 462 analyzed EMP stars are classified as mono-enriched. This means that the majority of EMP stars are likely multi-enriched, suggesting that the first stars were born in small clusters. Lower metallicity stars are more likely to be enriched by a single supernova, most of which have high carbon enhancement. We also find that Fe, Mg. Ca, and C are the most informative elements for this classification. In addition, oxygen is very informative despite its low observability. Our data-driven method sheds a new light on solving the mystery of the first stars from the complex data set of Galactic archaeology surveys.
\end{abstract}


\section{Introduction}
\label{sec:intro}
Big Bang nucleosynthesis has produced hydrogen, helium, and trace amounts of lithium. All heavier elements were synthesized and released by stars and their violent final fates, such as supernova (SN) explosions. The crucial transition from a primordial Universe to a Universe enriched with heavier elements (summarized as ``metals'' by astronomers) was initiated by the first stars, also termed Population~III (Pop~III) stars. These stars formed in pristine minihalos around redshift 6-30 \citep{bromm02,yoshida03,magg16,jaacks19,skinner20,schauer20,kulkarni20,liu20b,hartwig22}. They ended the cosmic dark ages, provided the first metals, contributed to the reionisation of the Universe, they might have provided the seeds for the first supermassive black holes \citep{woods19}, and they have set the scene for all subsequent galaxy formation \citep{tay14,dayal18,chen20,washinoue21}. Despite their importance for cosmology and intensive studies in the last decades \citep{glover05,greif15,klessen19,hammerle20}, only little is known about the first stars.

Pop~III stars are believed to be more massive than present-day stars because primordial gas cools slower than metal-enriched gas and the resulting scale for fragmentation is therefore larger \citep{bromm03}. This theoretical consideration on the masses of Pop~III stars is supported by the lack of observations of surviving metal-free stars. If Pop~III stars were less massive than $\sim 0.8 \Msun$, their lifetime should be longer than the age of the Universe and we could observe metal-free stars in the Milky Way. Based on their non-detection, the lower mass limit of the initial mass function (IMF) of the first stars is $\gtrsim 0.65 \Msun$ with 95\% confidence \citep{hartwig15,ishiyama16,magg18,magg19,rossi21}. Further observational constraints based on gravitational waves \citep{kinugawa14,hartwig16,liu20a}, high-z SN explosions \citep{hummel12,hartwig18b,rydberg20,regos20}, or direct detection \citep{schauer20b,Grisdale21,riaz22} may become available in the future.

Numerical simulations of Pop~III star formation are an important addition to deepen our understanding of primordial star formation. All recent simulations that resolve the protostellar disk show fragmentation of the primordial gas and predict that the first stars should form in small clusters \citep{clark11,greif15,liao19,wollenberg20,sharda20,sugimura20}. \citet{susa19} has shown that the number of fragments in a Pop~III-forming minihalo increases with time after the formation of the first protostar, and concluded that we should expect 10-50 Pop~III protostars per minihalo. However, no numerical approach has simulated the formation process until radiative feedback halts accretion onto the protostars \citep{hosokawa16} and the zero age main sequence (ZAMS) of Pop~III stars is reached. Therefore, we cannot make concluding statements about the final masses and multiplicity of Pop~III stars, because the gas accretion, stellar mergers, and possible ejections out of the minihalo are not yet fully simulated. In conclusion, all recent simulations suggest that Pop~III stars could form in a small cluster, but we do not yet have an observational confirmation for this consensual result of numerical simulations.

The multiplicity of the first stars is not only an academic curiosity, but has several direct implications. The Pop~III multiplicity defines the number of primordial stars per minihalo and combined with the characteristic mass of Pop~III stars, this quantity allows to constrain the star formation efficiency of primordial gas. The resulting cosmic number density of Pop~III stars also sets tighter constraints on their contribution to reionisation, to the Lyman-Werner background, and to the number of available stellar mass seed black holes. Moreover, stellar multiplicity is a necessary condition to form stellar binaries. Pop~III stellar binaries \citep{stacy13,sugimura20,liu21} might provide binary black holes that can be observed with gravitational wave detectors, or might evolve into X-ray binaries that can provide X-rays at high redshift or an additional source to cosmic reionization.

Another independent approach to indirectly constrain the nature of the first stars is stellar archaeology \citep[e.g.,][]{frebel15}. The first SNe have released metals into the previously pristine interstellar medium (ISM). Second-generation stars can form out of this enriched gas, which still carry the specific chemical fingerprint of the first SNe. Some of these second-generation stars have survived until the present day and we can observe them as old, metal-poor stars in the Milky Way. If we can determine the chemical composition of such extremely-metal poor (EMP, [Fe/H]\footnote{Defined as [X/H]$ = \log_{10}(N_\mathrm{X}/N_\mathrm{H})-\log_{10}(N_{\mathrm{X},\odot}/N_{\mathrm{H},\odot})$ throughout the paper, where $N_\mathrm{X}$ is the abundance of metals, $N_\mathrm{H}$ is the abundance of hydrogen, and $N_{\mathrm{X},\odot}$ and $N_{\mathrm{H},\odot}$ are the Solar abundances of these \citep{asplund09}.} $\leq -3$) stars with high-resolution spectroscopy, we can infer the nature of the first SNe and therefore constrain the properties of the first stars \citep[e.g.,][]{umeda03Nature,tominaga14,placco16,fraser17,ishigaki18,choplin19b,hansen20,skuladottir21,placco21,hartwig22}.

Previously, a standard assumption of stellar archaeology is that EMP stars are mono-enriched, i.e., that one EMP star contains metals from only one enriching SN \citep{Audouze95}. Based on this assumption, the initial mass function of the first SNe was observationally estimated \citep{ishigaki18}. 
However, recent numerical simulations and semi-analytical models \citep{hartwig18a,hartwig19b} suggest that Pop~III stars form in small clusters and that there should be multiple SN explosions in one minihalo. Consequently, we expect that at least some of the observed EMP stars are multi-enriched. While it is in principle possible to assume multi-enrichment in traditional abundance fitting methods, it increases the number of free parameters to fit the chemical composition of an EMP star and therefore weakens the predictive power \citep{chan17,salvadori19}.

In this study, we develop a new method with supervised machine learning (\S 2) to constrain the number of enriching Pop~III SNe, and apply the method to 462 EMP stars from literature. Our method allows us to constrain the multiplicity of the first stars, for the first time (\S 3). \S 4 and \S 5 give discussion and conclusions.

\section{Methodology}
This is the first approach to classify a representative set of EMP stars into mono- and multi-enriched. Therefore, there is no convenient ground truth for this problem, i.e., a catalogue that labels observed EMP stars with the number of SNe that enriched the gas out of which they formed. In order to deconvolve the physical properties of the first stars from complex observational data, we therefore took the following steps:
i) we carefully selected a sample of EMP stars;
ii) we then created a realistic subset of synthetic SN yields,
iii) which we then used to generate representative mock observations to
iv) train and optimize our Support Vector Machine (SVM) classifier.

\subsection{EMP Star Observations}
\label{sec:EMPobs}
In this study, we analyzed EMP stars with [Fe/H] $\leq -3$. We base this metallicity threshold on two main criteria: first, this is the conventional threshold to define EMP stars \citep{beers05}. Second, the metallicity contribution from Pop~III SNe dominates the overall metal mass up to [Fe/H] $\sim -3$ \citep{hartwig18a,ishigaki21}, which implies that EMP stars are mainly enriched by the first stars. We selected 432 EMP stars from the SAGA database \citep{saga}, added 26 stars \citep{yong13,cohen13,roederer14} from the compilation by \citet{ishigaki18} that have not yet been listed in the SAGA database, and added four recent stars at the lowest metallicities \citep{aguado18,nordlander19,skuladottir21,placco21}.

Our analysis is based on abundance ratios of metals with respect to each other and not with respect to hydrogen. This has two main reasons: first, we want to analyze the observed stars independently of their absolute metallicity. This enables us to extend the analysis in the future to stars with [Fe/H] $> -3$, where we also expect interesting, multi-enriched stars with contributions from Pop~III SNe \citep{salvadori19,ishigaki21,aguado23}. Second, we created mock observations to mimic the observed stars. Modeling the mixing of metals with pristine gas after Pop~III SNe is a stochastic process \citep{tarumi20} that requires high-resolution hydrodynamical simulations \citep{ritter15, sluder16} for every single star, which is beyond the scope of current computational capacities.

The data quality and availability of various elements differ among EMP stars, depending on, e.g., the instrument, signal-to-noise ratio, and covered spectral range. We created the mock observations to have the same observational properties as the sample of observed EMP stars that we analyzed. Moreover, we only included stars with a detected [Fe/H] value (and not upper limits) to investigate trends with metallicity. For all possible combinations of abundance ratios with elements between carbon and zinc, we calculate their observability as fraction of the EMP star sample for which this abundance ratio is measured. For example, the abundance ratio [Ca/Fe] is available for 452/462 EMP stars in our sample and [Ca/Mn] is still available for $\sim 63\%$ of EMP stars. Note that upper detection limits were counted as non-detections since they are not useful for our supervised classification. We include the following 13 elements, as we will motivate below: C, O, Na, Mg, Al, Si, Ca, Cr, Mn, Fe, Co, Ni, Zn.

These observed abundances reflect the current composition of the star in the stellar photosphere. However, the surface carbon abundance of a star can be reduced due to CN processing in the upper red giant branch. Hence, the observed carbon abundance can be smaller than the natal carbon abundance, which is relevant for this study. To correct for this effect, we apply the carbon corrections based on \citet{placco14}, which take into account the metallicity and surface gravity\footnote{\url{http://vplacco.pythonanywhere.com/}}. While this correction can affect the classification of individual EMP stars, it does not affect the mean fraction of multi-enriched EMP stars in our sample. Throughout the paper, we will use the corrected carbon abundances and only in App.~\ref{sec:Ccorr}, we show results without the carbon corrections.

\subsection{Theoretical Supernova Models}
We used of a set of Pop~III SN yields \citep{ishigaki18}, which are based on pre-SN and explosive nucleosynthesis calculations \citep{umeda00,umeda05,tominaga07}. Both mass cut and mixing are treated as free parameters under the framework of the mixing-fallback model \citep{umeda02,umeda05, tominaga07}. The mass cut is constrained empirically to include cases where the explosion is not spherically symmetric, in which case a single value of the mass cut may not be well defined \citep{umeda02,tominaga09,ezzeddine19}. These yields have been proven to be successful in reproducing the chemical composition of individual EMP stars, including carbon-enhanced metal-poor (CEMP) stars \citep{tominaga07,ishigaki18}, as well as to explain the more complex chemical evolution of the Milky Way \citep{kobayashi06} and Damped Lyman-$\alpha$ (DLA) Systems \citep{kobayashi11}.

To ensure that we only took into account yields of realistic Pop~III SNe, i.e., SNe that occur in nature, we determined how successful a particular SN yield model is at reproducing observed abundance patterns of EMP stars. We did this by calculating the reduced $\chi ^2$ between each model and the selection of EMP stars \citep{ishigaki18} (differences between the [X/H]$_i$ values), taking into account element-specific observational uncertainties. For every yield model, we noted the mean $\chi ^2$ that it can provide, averaged over all observations.

The goal was to identify a threshold so that all yield models with $\chi ^2$ below this value can be considered realistic SN yields. To find such an an optimal, data-driven threshold, we compared the distributions of the 78 considered abundance ratios between the selected subsets (sets of SN yield models with $\chi ^2$ below a certain threshold) and our set of 462 observed EMP stars. We found that the range $\chi ^2<15.1$ provides the best agreement between the abundance ratios of observed EMP stars and Pop~III SN yields (see Sec.~\ref{sec:VaryYields} for details). This threshold on $\chi ^2$ selects 512 different SN yield models.

We also added 27 yield models that were identified as best fits to individual EMP stars \citep{ishigaki18} and that were not yet included. While the first subset accounts for yields that are a good average fit to many EMP stars, the second subset guarantees that more exotic yields that are a good fit to only one or few EMP stars are included. These additional 27 yield models include abundances with low [Mg/Fe] and high [C/Fe].

Although we select a theoretical yield set under the assumption of mono-enrichment to determine the best fits, the yield set is representative of both, mono- and multi-enrichment. This is because multi-enriched SN yields usually form a centrally concentrated subset in the abundance space of mono-enriched yields. With this approach, we can reject the null hypothesis that all EMP stars are mono-enriched.

Our subset of theoretical yields comprises 539 SN yield models, including progenitor stars of masses (13, 15, 25, 40, 100)\Msun\, and explosion energies of $(0.5,1,10,30,60) \times 10^{51}$\,erg \citep{ishigaki18}. Most of the selected SN models are normal CCSNe with progenitor masses in the range $13-25\Msun$ or hypernovae with $25-40\Msun$, and 24\% of our training set comprises faint SNe with ejected mass of $^{56}$Ni $\lesssim 10^{-3}$\Msun, which decays to Fe.

\subsection{Mock Observations}
\label{sec:MockObs}
In this section, we explain the main formalism to discriminate enrichment by one vs. multiple SNe. To create mock observations of mono-enriched stars, we directly took the masses of yields from the selected 539 Pop~III models.

We created mock observations of multi-enriched stars by adding the yields of multiple randomly selected SN yield models together. Here, we implicitly assume that the metals from enriching SNe mix homogeneously. While this assumption may not be valid under all conditions \citep{ritter15,sluder16}, it avoids the inclusion of unconstrained mixing fractions. We did not take into account the effects of inhomogeneous mixing between different elements \citep{chiaki20} or the energy-limited hydrogen dilution mass \citep{magg20}.

Multi-enrichment could result from any combination of $\ge 2$ SNe. In a pilot study, we attempted to discriminate between different levels of multi-enrichment from 2, 3, 4, or 5 SNe. However, the discrimination of N-fold enrichment is very degenerate, and it is already challenging to discriminate enrichment from one and two SNe. Therefore, we only distinguished between mono- and multi-enrichment in this study and did not aim to quantify the level of multi-enrichment. First, this makes our prediction more robust because we mitigate the degeneracy for various levels of multi-enrichment. Second, we are mainly interested in discriminating mono- and multi-enrichment and the exact level of multi-enrichment is only of secondary importance. Therefore, we created multi-enriched mock observations by combining two-fold, three-fold, four-fold, and five-fold enrichment into one set with the same size as the set of mono-enriched mock observations. In addition, we verified that the exact number and combination of SNe that we include in the multi-enriched set does not affect the final classification of mono- or multi-enrichment.

We augmented the amount of multi-enriched mock observations by a factor 16 to include additional possible combinations of multi-enrichment. With four different levels of multi-enrichment (enrichment from 2-5 SNe) and a training set size of 50\%, this provided us with approximately 17,000 multi-enriched mock observations for training and we generate the same amount of mono-enriched mock observations. In all our sets for training, cross-validation, and blind test, the fraction of mono- and multi-enriched samples is always 50\% each. In other words, our prior assumption is that mono- and multi-enrichment are equally likely (see Sec.~\ref{sec:prior} for more details).

Once we calculated the abundance ratios of our mock observations, we applied the observational masks to mimic the observability: for each of our 462 EMP stars, we determined the observability masks, i.e., an array that contains the information if a certain abundance ratio is observable. We then applied these masks to our validation and test data. This guaranteed that we used the same information content when analyzing the mock observations, as it is available when we eventually evaluate the actual EMP stars.

In this approach, we began with 13 elements C, O, Na, Mg, Al, Si, Ca, Cr, Mn, Fe, Co, Ni, and Zn, which enables the construction of 78 independent abundance ratios. We wanted to keep the number of elements small since the complexity of the model scales with the number of elements and more complex models are prone to overfitting.
Therefore, we only used elements for which we have a sufficient number of observations for EMP stars. We also excluded Sc and Ti because our theoretical models cannot correctly resolve the production of these elements \citep{ishigaki18,tominaga09,kobayashi20}.

While N can also provide constraints on Pop~III SNe, theoretical uncertainties of N yields are large due to the potential presence of stellar rotation \citep{meynet02,hirschi07,choplin19}. Since the number of N measurements is also small, we do not use N in this work.

Finally, to account for theoretical uncertainties, we added random scatter to the mock observations. We introduce a matrix of error bars for specific combinations of abundance ratios because some of the elements share the same physical or observational reasons for uncertainties (see App.~\ref{sec:uncert}). Note that we include observational uncertainties at classification stage with Bootstrap sampling (see below).

\subsection{Ground Truth Sample}
Fig.~\ref{fig:SNMonoMulti} presents the distribution of mock observations enriched by one (orange circles) or multiple (blue triangles) SNe.
\begin{figure}
	\includegraphics[width=\columnwidth]{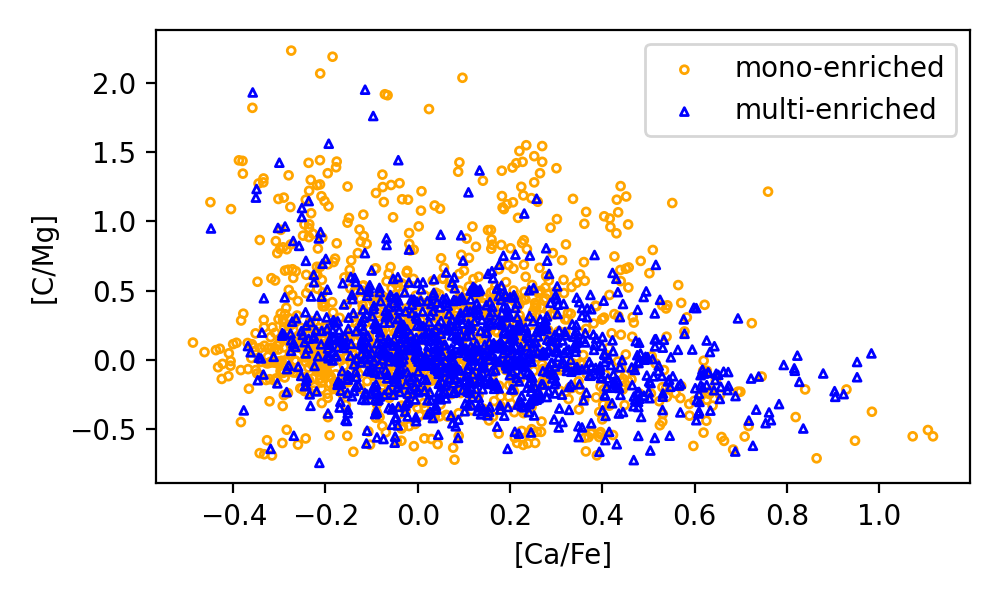}
    \caption{Ground truth training data for EMP stars enriched by one (orange circles) or multiple (blue triangles) SNe. Multi-enriched EMP stars are more centrally concentrated because their yields are a weighted average of individual SNe. Exceptions to this trend can result from theoretical uncertainties that are added as scatter.}
    \label{fig:SNMonoMulti}
\end{figure}
In this example, we used [C/Mg] and [Ca/Fe] as the two dimension since we show below that these are the most informative elements for this classification purpose. In general, mono-enriched stars span a larger area in this abundance space, whereas multi-enriched stars are more clustered towards the center. This makes sense because the abundance ratio of a multi-enriched star is the weighted mean of the individual abundance ratios of the individual SNe that contributed to the enrichment. In other words, by combining two SN yields, the resulting abundance ratios will never be more extreme than the abundance ratios of any of the individual contributing SNe. This figure also illustrates a degeneracy where the abundance ranges are largely overlapping if only a small number of elements are used. As a consequence of this degeneracy, mono-enriched stars can be classified more reliably because there are regions of the parameter space that can only be reached by the yields of single SNe. Hence, we mostly focus on the probabilities and fractions for mono-enrichment, as they are more reliable.

\subsection{Supervised Classification with Support Vector Machines}
The basis of this approach are SVMs, a supervised machine learning technique that iteratively finds a hyperplane in the feature space that optimally discriminates two classes \citep{SVM95}. SVMs have been applied successfully to various astrophysical tasks \citep{SVM2,SVM1,SVM5,SVM3}. If the data is not linearly separable (like in our case), the SVM attempts to find an optimum by minimizing the number of misclassifications and their distance to the decision boundary.

We show below that four dimensions is optimal for the SVMs in this problem. Hence, the training data are points $x_i \in \mathbb{R}^4$ with their associated classes $y_i \in \{-1,1\}$. The learning goal for the linear SVM is to find optimal $w, b, \zeta$ that minimize
\begin{equation}
    \frac{1}{2}w^Tw + C \sum _i \zeta _i
\end{equation}
under the constraint
\begin{equation}
    y_i(w^Tx_i + b) \ge 1 - \zeta _i,
\end{equation}
with $\zeta _i > 0 $ and the index $i$ running over all training examples \citep{SVM11}. The second equation guarantees that every point is on the correct side of the decision boundary or, at most, $\zeta _i$ away from their correct margin boundary. The first equation aims at maximizing the margin between the two classes by minimizing $||w||$. The second term minimizes the allowed tolerance $\zeta _i$ with the regularization parameter, for which we find $C=1$ to be an optimal choice. The final classification of the SVM for a new observation $x_i$ is obtained by evaluating the sign of $w^Tx_i + b$.

Instead of using the points $x_i$ directly in the linear SVM, one can also augment the data or replace the $x_i$ with a kernel function $\phi (x_i)$. This Kernel Trick might enable the transformation of training data that is not linearly separable in 4D into linearly separable data in higher dimensions. We found that radial basis functions with a kernel width of $\gamma=1$ are an ideal choice for our specific classification problem.

For the training, we assumed that all abundance ratios are observable. For the cross-validation and blind test data, we accounted for the fact that certain abundances and combinations of abundance ratios are not always observable in all EMP stars.

\subsection{Ensemble Learning}
A single supervised classifier is susceptible to overfitting. Consequently, a single SVM has a higher generalisation error and might perform poorly on new, unseen data. To overcome this problem, we employed Ensemble Learning and combined multiple SVMs into one final prediction.
One additional advantage of Ensemble Learning for our approach is that we can fully exploit the available abundance ratios.

The optimal dimension for the SVM is a balance between two factors: a higher dimension provides better classification accuracy because it is based on more information and the SVM has more flexibility to identify an optimal decision boundary. On the other hand, each dimension introduces one new abundance ratio, which might not be observable for EMP stars. If only one element is not available, this SVM cannot be used for such an observation. Thus, we find that 4D is a good compromise.

We combined multiple SVMs to avoid overfitting and mitigate missing data. The strategy was that we trained several independent SVMs and let them vote on the final classification. In addition, we performed Bootstrap sampling to mimic observational uncertainties. The final classification for one star is an average of up to $N_\mathrm{B} \times N_\mathrm{SVM}$ individual classifications, where $N_\mathrm{B}=49$ is the number of Bootstrap samples, and $N_\mathrm{SVM} \leq 10$ is the number of used SVMs in this paper.

We first resample the observed abundance ratios 49 times, based on the uncertainty matrix (see App.~\ref{sec:uncert}). For the number of resamplings, we have chosen an odd number to break ties. The results do not change with a larger number of resamplings. For each sample, we obtain a classification result ($ p_\mathrm{mono,i} \in [0,1]$, i.e., 0 for mono-enriched and 1 for multi-enriched in our case). For an individual EMP stars, we obtain its mean value and uncertainty over the bootstrap samples in the form $p_\mathrm{mono}=\bar{x} \pm \sigma _x$. For the entire sample of EMP stars, we first calculate the mean fraction of mono-enriched EMP stars as $f_\mathrm{mono,i}$ for one bootstrap sample. Then, we calculate the mean value and uncertainty of mono-enriched fraction for the entire sample based on the $N_\mathrm{B}$ bootstrap samples. Throughout the paper, we use this latter approach (resampling the ensemble of EMP stars) as our fiducial model and report the uncertainty of individual EMP stars wherever appropriate.

If the abundance ratio for some SVMs is not available, they cannot provide a prediction, and we take the average only from SVMs that can be used for that specific observation. Therefore, it is essential to select an optimal configuration of SVMs that fully exploits the available abundance ratios. With 78 different abundance ratios, there are over one million possible combinations to choose the four dimensions of an SVM. To find the optimal combination, we proceeded in the following manner: we created all possible quadruples of abundance ratios and sorted them by observability. Then, we selected the top 10 most observable abundance ratio quadruples and augmented them with 6 additional quadruples so that each of the 13 considered elements is included at least once in a quadruple. Next, we trained 16 SVMs with these most promising quadruples of abundance ratios as dimensions. This provided us with 16 different 4D SVMs that are trained on different abundance ratios. Then, we used the cross-validation set to test how all possible combinations of these 16 SVMs would perform.

We selected our optimal model based on the validation accuracy and several other constraints: we aimed for a maximum fraction of (mock) observations to be classified. Hence, we prefer combinations of abundance ratios with high observability. Moreover, we required a symmetric confusion matrix and a large dynamical range of possible predictions. Our optimal fiducial model is based on 10 different 4D SVMs that use 24 different abundance ratios. The two elements Mn and Ni are not included in the final set of 10 SVMs. This selection is data-driven, based on the input yields and our optimization goals.

We also considered other classification algorithms, such as deep neural networks, decision trees, and a regressive approach. Since we created our own mock observations for the training, our problem is not limited by data. In such a case, most classification algorithms show a similar marginal performance because with sufficient training data and a consequently dense training set, every classification problem becomes a nearest neighbour search. We verified this in the initial phase of the project and similar behaviour was shown by independent studies on astrophysical and generic data \citep{bazell01,lan20}. Moreover, SVMs produce inspectable models and, therefore, provide a certain intuition regarding how the decision boundary is derived. This is very important for the physical interpretation of the results and, therefore, provides an additional validation.

\subsection{Is the Mock Data Realistic?}
\label{sec:realistic}
The first important question is if our training data is representative of the actual data \citep{acquaviva2020debunking}. For example, if we train a supervised classifier to discriminate cats and dogs, but then use it to distinguish elephants and penguins, the final classification will not be accurate, even if the test accuracy on a set of cats and dogs is good.

\begin{figure}
	\includegraphics[width=\columnwidth]{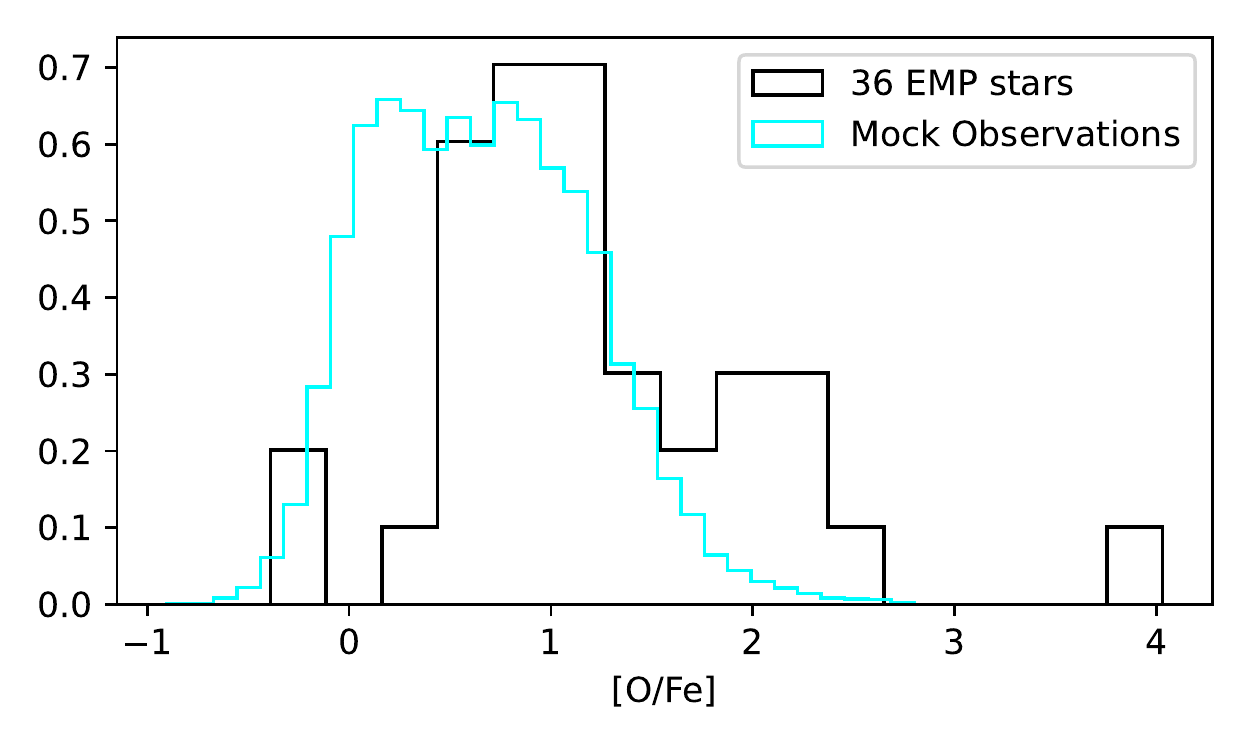}
	\includegraphics[width=\columnwidth]{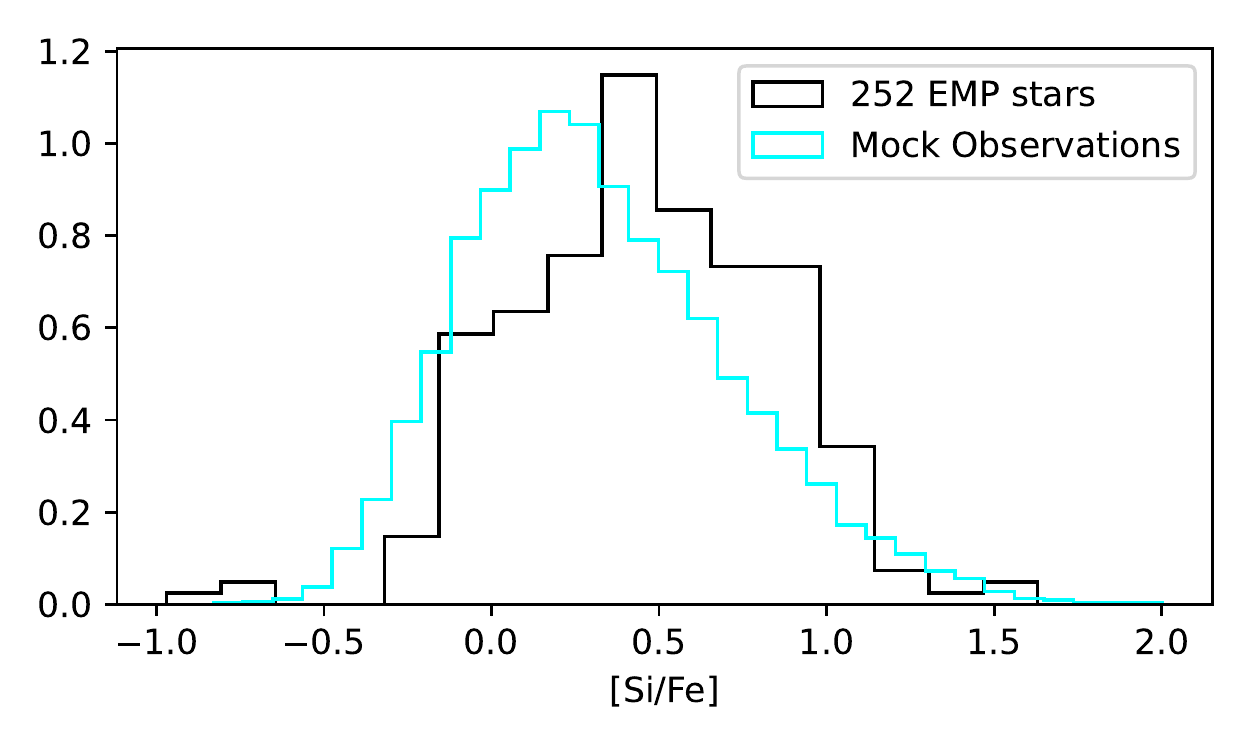}
	\includegraphics[width=\columnwidth]{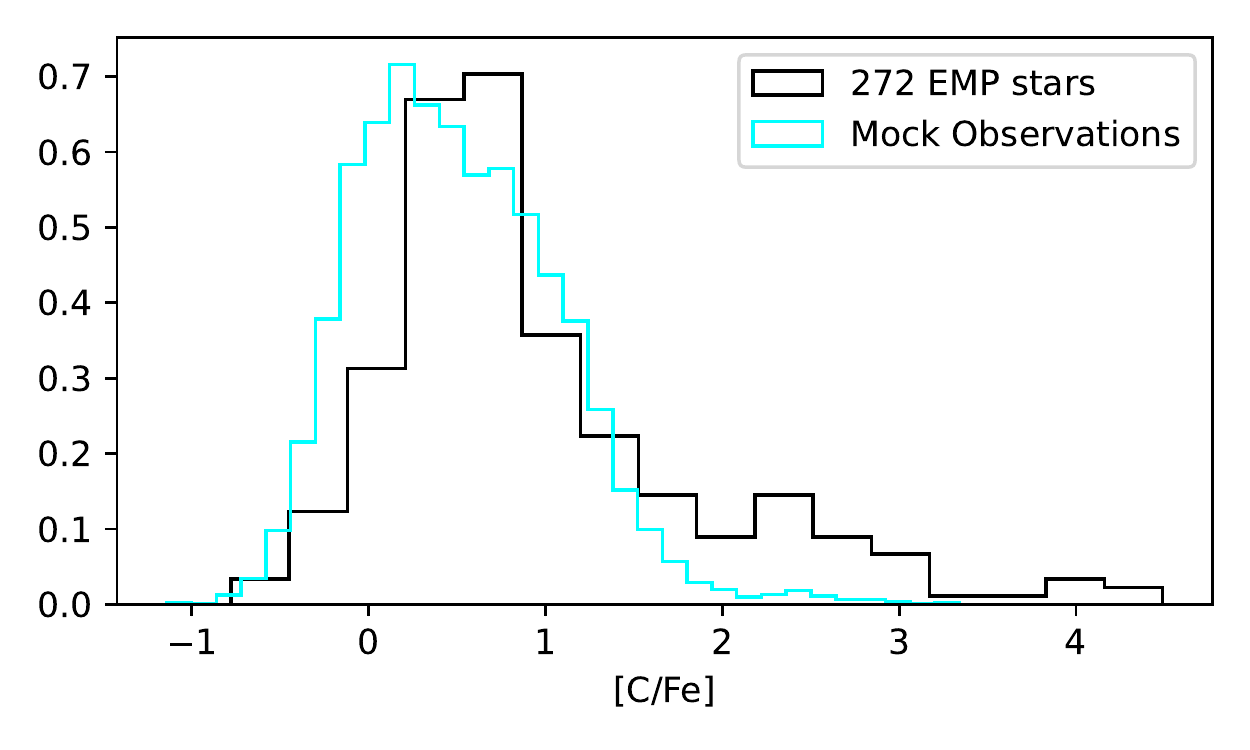}
    \caption{Comparison of the abundance distribution for the three most informative abundance ratios. While the distributions look qualitatively similar, the distribution of [O/Fe] is shifted by $>0.5$\,dex, which might introduce a bias in the final classification.}
    \label{fig:HistTrain}
\end{figure}

In Fig.~\ref{fig:HistTrain}, we compare the abundance distributions from the mock observations and the EMP stars for three representative abundance ratios. In the Appendix, we show the remaining 21 histograms for all abundance ratios used.
By construction, our subset of Pop~III SN yields minimizes the mean KS distance between mock observations and real EMP stars. Phrased differently, it is the best subset that we could select for this purpose. Here, we check if the best is good enough.

For most of the abundance ratios, the distributions of mock and actual observations are similar. However, there are discrepancies in the distributions for few abundance ratios, which may introduce a bias in our conclusions. There is no independent method to confirm if our mock observations are realistic. However, three reasons support our approach to constructing the mock observations.

First: We verified that our final conclusion is not affected when we modify the subset of Pop~III SNe that we use as training set (see Sec.~\ref{sec:VaryYields}). In summary, we increased and decreased the number of included Pop~III SN yields by a factor of two, which did not significantly change the final results. Moreover, we shifted and scaled the values of the 10 most informative abundance ratios so that the mean and standard deviation of their distributions match exactly with the distribution of the EMP stars. This did also not significantly affect the final classification. Therefore, our training data might not be perfect, but we are confident that the final conclusion of this study is not affected by the exact selection of training data.

Second: The final decision of the SVMs is based on a majority vote. If the combined classifications of the independent SVMs is within $1\sigma$ of the decision boundary, we can consider the sample is not classified. This makes the decision process more robust towards new, unseen observations. Thus, even if the ranges of mock and actual observations are not precisely overlapping, our classification process is sufficiently robust to handle outliers.

Third: If all these histograms for all abundance ratios matched exactly, we would not need this study. It would imply that we understand the enrichment and formation history of EMP stars sufficiently well. Unfortunately, this is not the case. The mock observations are constructed based on our current understanding of EMP enrichment. Therefore, the mismatch between actual observations and mock observations highlights that our current understanding is not sufficient. Instead, based on our current state-of-the-art assumptions regarding Pop~III SNe and EMP stars, we construct our mock observations and thereby contribute to iteratively improving our understanding of the physics involved.

\subsection{Accuracy of SVMs}
\label{sec:acc}
\begin{figure}
	\includegraphics[width=\columnwidth]{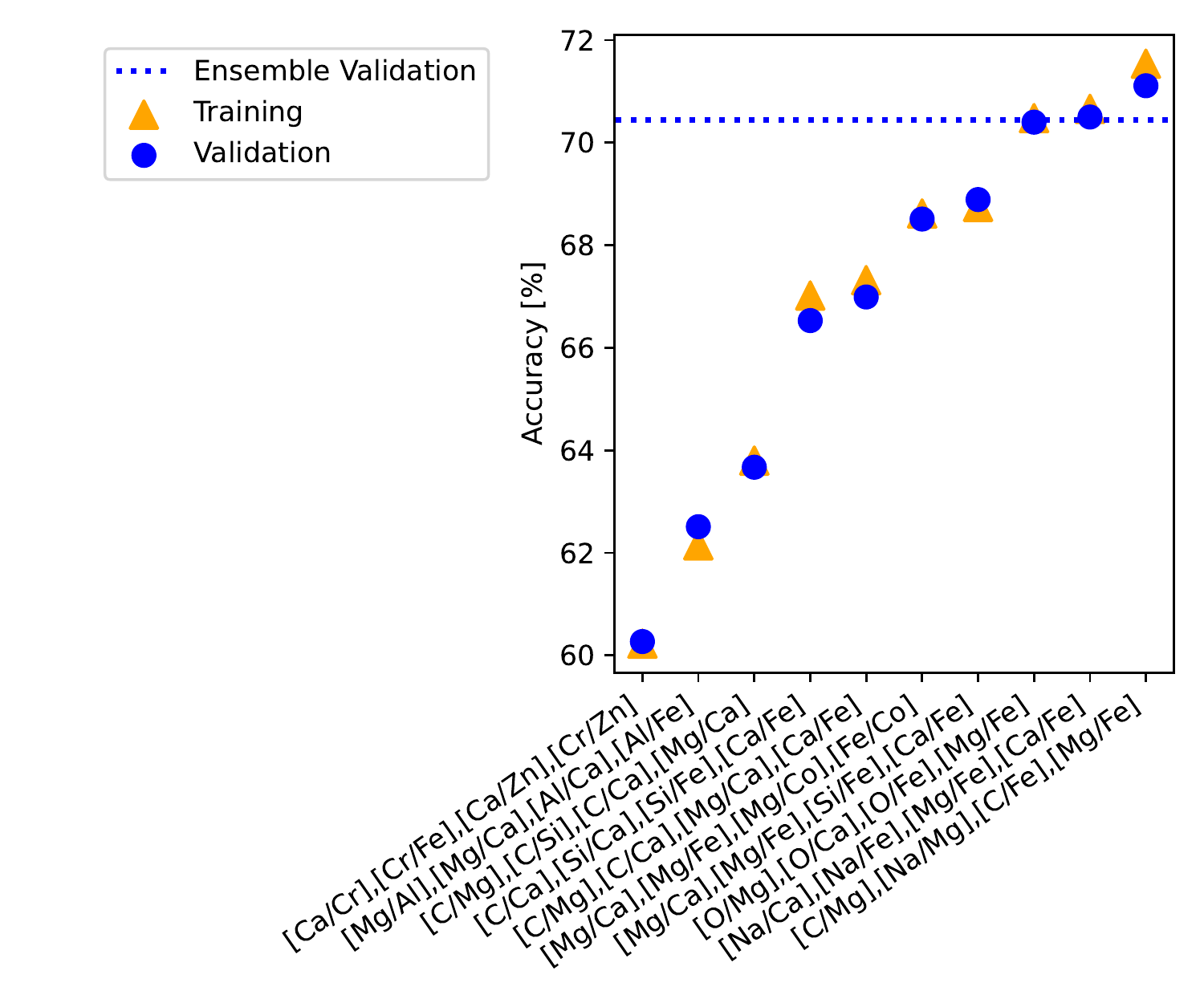}
    \caption{Prediction accuracy on the training set (triangles) and validation set (dots) for the individual SVMs. The dashed line shows the validation accuracy of the ensemble, i.e., the combined prediction of the 10 SVMs. Individual SVMs can achieve a higher accuracy than the ensemble method. However, such SVMs need specific elements (O, Na), which are not observable for all stars. The ensemble validation accuracy is valid for all classifications, irrespective of the number of observed elements. This high ensemble validation accuracy demonstrates the advantage of an ensemble learning approach. The order of the 10 SVMs from left to right is sorted by accuracy as visual aid.}
    \label{fig:acc}
\end{figure}

We measured the prediction accuracy on the training set and on a second validation set that was not used in the training. We defined the accuracy as the fraction of correctly classified mock observations. The accuracy for the individual SVMs and for the combined prediction is presented in Fig.~\ref{fig:acc}.
The validation accuracy provides an independent estimate on unseen data. The validation accuracy of the Ensemble Learning model is higher than most individual accuracies. This demonstrates the advantage of a weighted vote between the 10 SVMs compared to the prediction of a single SVM.

\begin{figure}
	\includegraphics[width=\columnwidth]{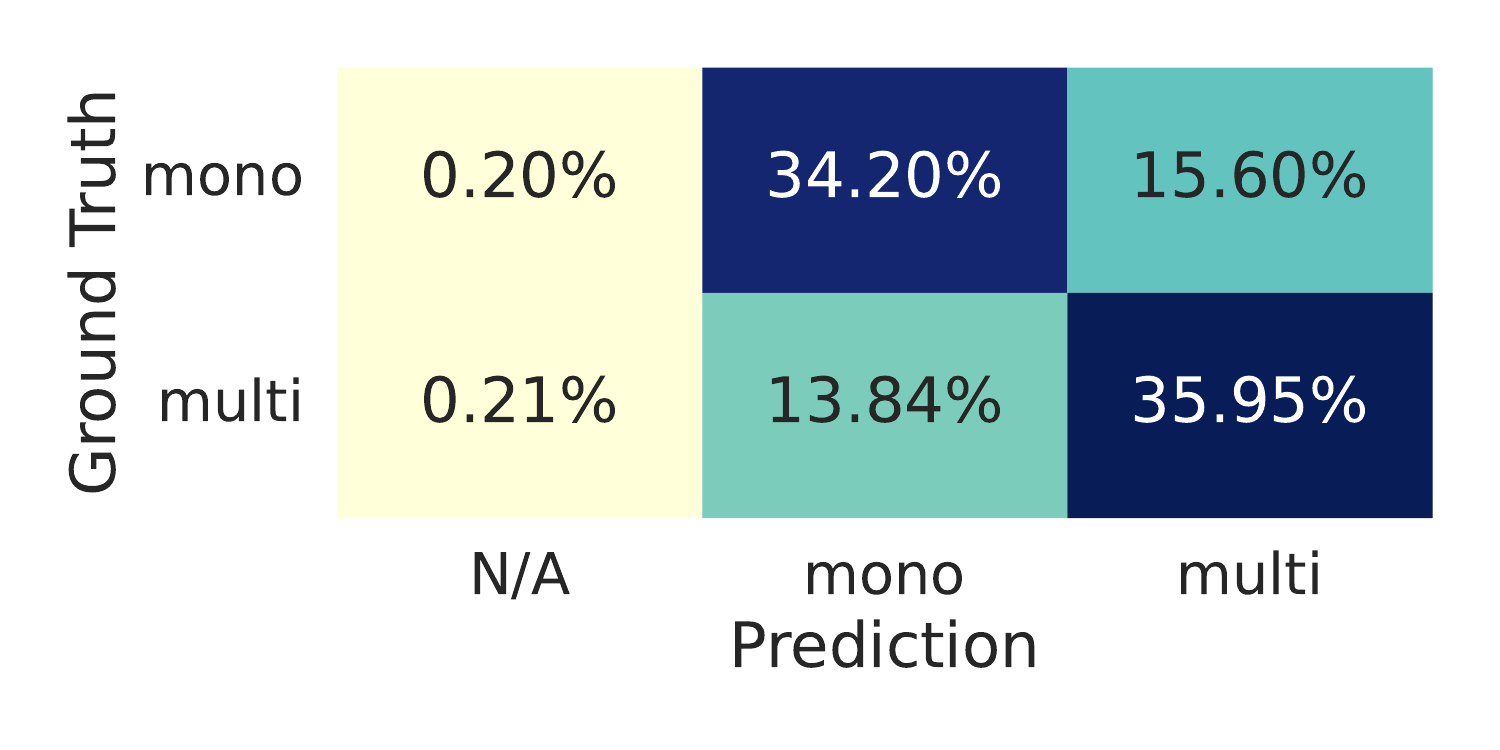}
    \caption{Confusion matrix of our Ensemble Classifier. The confusion matrix is balanced and symmetric.}
    \label{fig:conufs}
\end{figure}

To further understand the predictions, errors, and associated biases, we illustrate the confusion matrix in Fig.~\ref{fig:conufs}.
Overall, the confusion matrix is symmetric: the misclassifications are almost equally distributed with $\sim 13.8\%$ and $\sim 15.6\%$ and the correctly classified cases are also almost equal with $\sim 34.2\%$ and $\sim 36.0\%$. Our prediction pipeline predicts mono-enrichment for $\sim 48.0\%$ and multi-enrichment for $\sim 51.6\%$ of the classified cases.

To further quantify the Bayes error and degeneracy of this problem, we repeated the training of our model, excluding theoretical or observational uncertainty. Under these optimal conditions, we found $79\%$ of mock observations are correctly classified (compared to $70\%$ with realistic uncertainties). Phrased differently, the intrinsic degeneracy of this classification task is a major challenge, and improved observational data can increase the accuracy, but will not be able to solve this underlying degeneracy completely.

\section{Results}
\begin{figure}
	\includegraphics[width=\columnwidth]{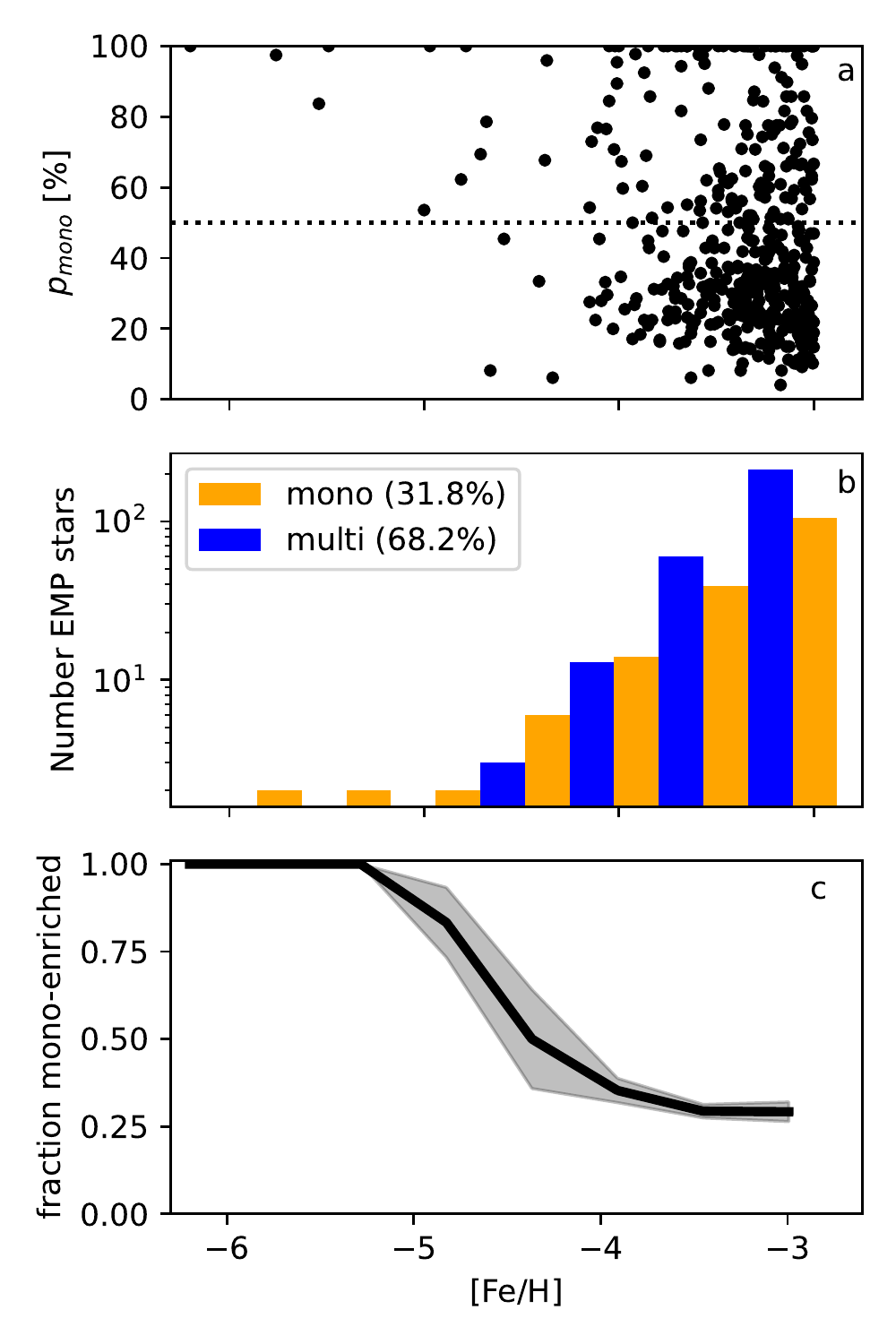}
    \caption{All panels show the results as a function of metallicity. Panel (a) shows the probability of mono-enrichment, $p_\mathrm{mono}$, for individual EMP stars. Panel (b) shows a histogram with the absolute number of classified mono-enriched (orange) and multi-enriched (blue) EMP stars in each bin (bins are slightly offset for clarity). Panel (c) shows the fraction of mono-enriched EMP stars, and the grey contour illustrates scatter due to the bootstrap resampling of observational uncertainties.}
    \label{fig:NSNFeH}
\end{figure}

In this paper, we analyze whether EMP stars are likely to be enriched by a single SN, or by multiple SNe in order to explain the observed elemental abundances.
For this purpose, we calculated nucleosynthesis yields of over 13,000 SNe covering all possible parameters for the first stars. Then, we trained an ensemble of SVMs on mock observations to classify high-resolution spectroscopic data of EMP stars ([Fe/H] $\leq -3$), using detailed chemical compositions from carbon to zinc assuming an equal contribution of  mono- and multi-enrichment in our training sample. The possibility of mono enrichment, which is determined by 10 SVMs and 49 Bootstrap samplings, is shown in Fig.~\ref{fig:NSNFeH} Panel (a).
From the original 462 EMP stars, we exclude 35 stars because they have only three elements observed, which is not enough for a reliable classification. Moreover, three EMP stars have a predicted $p_\mathrm{mono}=0.5$ and can therefore not be assigned to any category. From the remaining 424 EMP stars, we find the average fraction of mono-enriched stars to be $31.8\% \pm 2.3\%$, where the standard deviation reflects observational uncertainties.

As the number of stars sharply decrease toward lower metallicties, both numbers of  mono- and multi-enriched stars in our sample decrease at lower metallicities in panel (b). Panel (c) indicates a clear metallicity dependence of the ratio between them; stars at the lowest metallicities are likely to be mono-enriched, which has been assumed in numerous previous studies \citep{umeda03Nature,placco15,ishigaki18} but never been tested. At higher metallicities, stars tend to be multi-enriched. Our observation-based confirmation of this trend is remarkable, since we do not use [Fe/H] values to train our SVMs but use solely relative abundance ratios of various metals, excluding hydrogen. What is surprising is that the mono-enriched fraction is not 100\% at [Fe/H] $\sim -4.5$, which means that some second-generation stars were already enriched by multiple SN explosions.

\subsection{CEMP stars}
\begin{figure*}
\centering
	\includegraphics[width=0.8\textwidth]{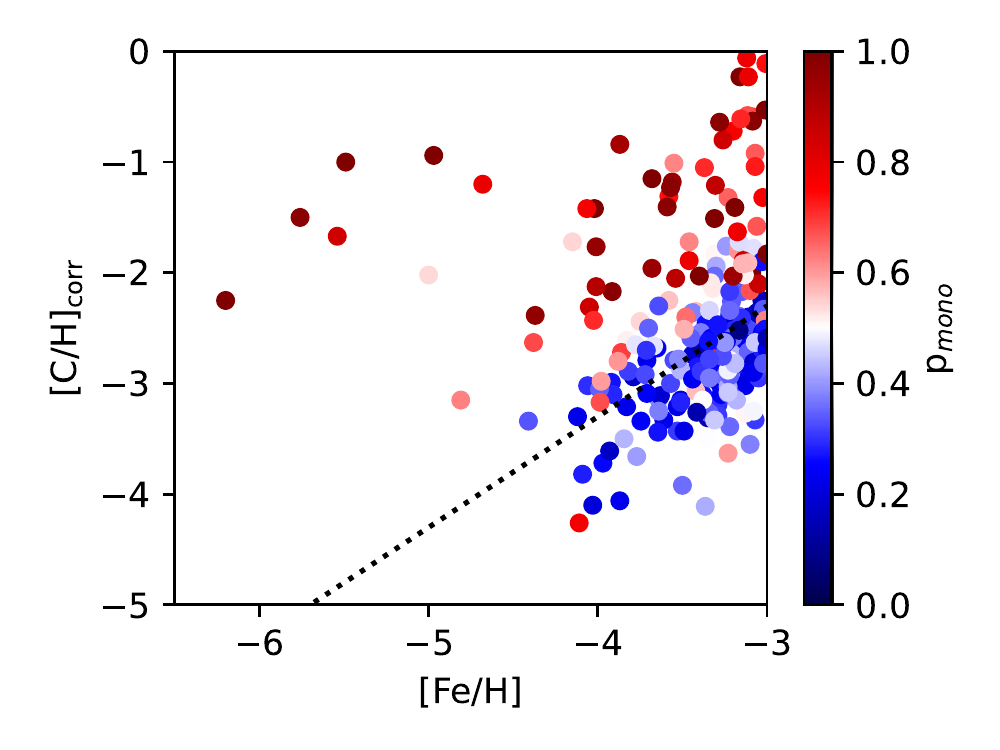}
    \caption{Carbon vs. iron abundance of EMP stars. The colour bar shows the probability for mono-enrichment. The dashed line at [C/Fe] $=0.7$ should guide the eye to highlight the range of CEMP stars. There is a trend that most CEMP stars are mono-enriched.}
    \label{fig:CFe}
\end{figure*}

For EMP stars, the most notable feature is their carbon enhancement; a large fraction of EMP stars show a large carbon enhancement relative to iron \citep{beers05,placco14} and multiple populations have been identified in the diagram of [C/H]--[Fe/H] \citep{bonifacio15,yoon16}. In Figure \ref{fig:CFe}, we depict the carbon vs. iron abundance of EMP stars, colour-coded by the probability for mono-enrichment.
We find a positive correlation between $p_\mathrm{mono}$ and [C/Fe]. EMP stars with high [C/Fe] are less likely to be multi-enriched. Specifically, 75 of 125 carbon-enhanced metal-poor stars (CEMP, [C/Fe] $>0.7$ \citealt{aoki07,arentsen22}) are mono-enriched, and all 49 stars with [C/Fe] $>1.5$ are mono-enriched. In the terms used in previous work \citep{spite13,bonifacio15,yoon16}, all Group~III stars are mono-enriched. The origin of this bi- or multi-modality can be explained as follows:
Multi-enrichment tends to average yields and makes them more centrally concentrated in the abundance space. Faint SNe, which are known to be important in the early Universe \citep{umeda03Nature,kobayashi11}, produce only small amounts of iron due to their larger black hole than for normal SNe; mixing their yields with normal CCSNe attenuates the initially high [C/Fe] from the faint SN to a smaller value. 
Therefore, it becomes more difficult for CEMP stars to form after multiple SNe have exploded in one minihalo \citep{jeon21}.
On the other hand, for C-normal stars, once they are enriched by a normal SN, it becomes impossible to eliminate the possibility of additional enrichment from faint SNe in our analysis; hence our estimated number of SNe is a lower limit.

One could speculate if mono-enrichment and carbon enhancement are synonyms, or one is a subset of the other. To allow readers their own conclusion based on the preferred threshold of carbon enhancement, we provide the quantitative classification data in the carbon-enhanced regime in Table~\ref{tab:CEMP}.
\begin{table}
\centering
\caption{Classification results in the carbon-enhanced regime as a function of the [C/Fe] threshold. The last column shows the fraction of all EMP stars for which mono-enrichment and carbon enhancement (based on the variable threshold in the first column) are synonym. At [C/Fe] $>1.5$ all CEMP stars are mono-enriched.}
\label{tab:CEMP}
\begin{tabular}{cccc} 
	\hline
	[C/Fe]$_\mathrm{corr}$ & N$_\mathrm{mono}$ & N$_\mathrm{multi}$ & CEMP $\Leftrightarrow$ mono-enriched\\
	\hline
$>0.7$ & 75 & 50 & $74.1\%$\\
$>0.8$ & 73 & 35 & $77.6\%$\\
$>0.9$ & 74 & 24 & $80.0\%$\\
$>1.0$ & 70 & 16 & $80.9\%$\\
$>1.1$ & 66 & 9 & $81.6\%$\\
$>1.2$ & 61 & 6 & $81.1\%$\\
$>1.3$ & 56 & 5 & $80.2\%$\\
$>1.4$ & 52 & 2 & $80.0\%$\\
$>1.5$ & 49 & 0 & $79.9\%$\\
	\hline
\end{tabular}
\end{table}

The fraction of mono-enriched stars increases with [C/Fe]. The last column shows the fraction of all EMP stars for which carbon-enhancement is a consequence of being mono-enriched \textit{and} mono-enrichment is a consequence of being carbon enhanced. I.e., the missing stars to 100\% are those that are either mono-enriched but not carbon enhanced, or that are carbon enhanced but not mono-enriched. This fraction is highest around [C/Fe] $\sim 1.1$. It declines at higher [C/Fe] because there are too many mono-enriched stars, which are not classified as carbon enhanced any more due to the higher threshold. Phrased differently, if we want to define a physics-informed threshold for CEMP stars based on the ability to discriminate mono- from multi-enriched EMP stars, the best threshold would be around [C/Fe] $\sim 1.1$.

The classification of EMP stars and their distribution on the [C/H]-[Fe/H] is affected by the carbon corrections. Therefore, we also provide a version of this figure without the carbon corrections in App.~\ref{sec:Ccorr}.

\subsection{Most metal-poor stars}
In Tab.~\ref{tab:HMP}, we show the classification of the most iron-poor stars in our sample.
\begin{table*}
\centering
\caption{List of the 10 most iron-poor stars in our sample. The columns show the name, iron abundance, carbon-to-iron ratio, number of available abundance ratios, and $p_\mathrm{mono}$ as result of our supervised classification. The full table is available online.}
\label{tab:HMP}
\begin{tabular}{lllll} 
	\hline
	Name & [Fe/H] & [C/Fe]$_\mathrm{corr}$ & $N_\mathrm{avail}$ & $p_\mathrm{mono}$ [\%]\\
	\hline
SMSS J160540.18-144323.1 &	$-6.20$& 3.95&	15&	$100$\\
HE 1327-2326 &	$-5.76$& 4.26&	15&	$97 ^{+3} _{-8}$\\
HE 0107-5240 &	$-5.54$& 3.87&	10&	$84 \pm 16$\\
SDSS J081554.26+472947.5&	$-5.49$& 4.49&	15&	$100$\\
SDSS J131326.89-001941.4 &	$-5.00$& 2.98&	21&	$54 \pm 9$\\
SDSS J092912.33+023817.0  &	$-4.97$& 4.03&	6&	$100$\\
HE 0557-4840 &	$-4.81$& $1.66$&	28&	$62 \pm 14$\\
SDSS J174259.67+253135.8 &	$-4.79$& 3.62 &	3&	$100$\\
SDSS J102915.14+172927.9 &	$-4.71$& -- &	10&	$69 ^{+31} _{-46}$\\
HE 0233-0343 &	$-4.68$& $3.48$&	10&	$79 \pm 9$\\
	\hline
\end{tabular}
\end{table*}
It will be interesting to model their exact formation scenarios based on the number of enriching SNe in future works. However, one has to be cautious with the direct interpretation of the provided face values. About $70\%$ of samples in the blind test set have been classified correctly ($80\%$ if we only take into account EMP stars for which the predicted $p_\mathrm{mono}$ is more than one standard deviation away from the decision boundary). Therefore, while the average fraction of mono-enriched stars is reliable, individual values for the number of enriching SNe should not be over-interpreted.

\subsection{Most informative elements}
\begin{figure}
	\includegraphics[width=\columnwidth]{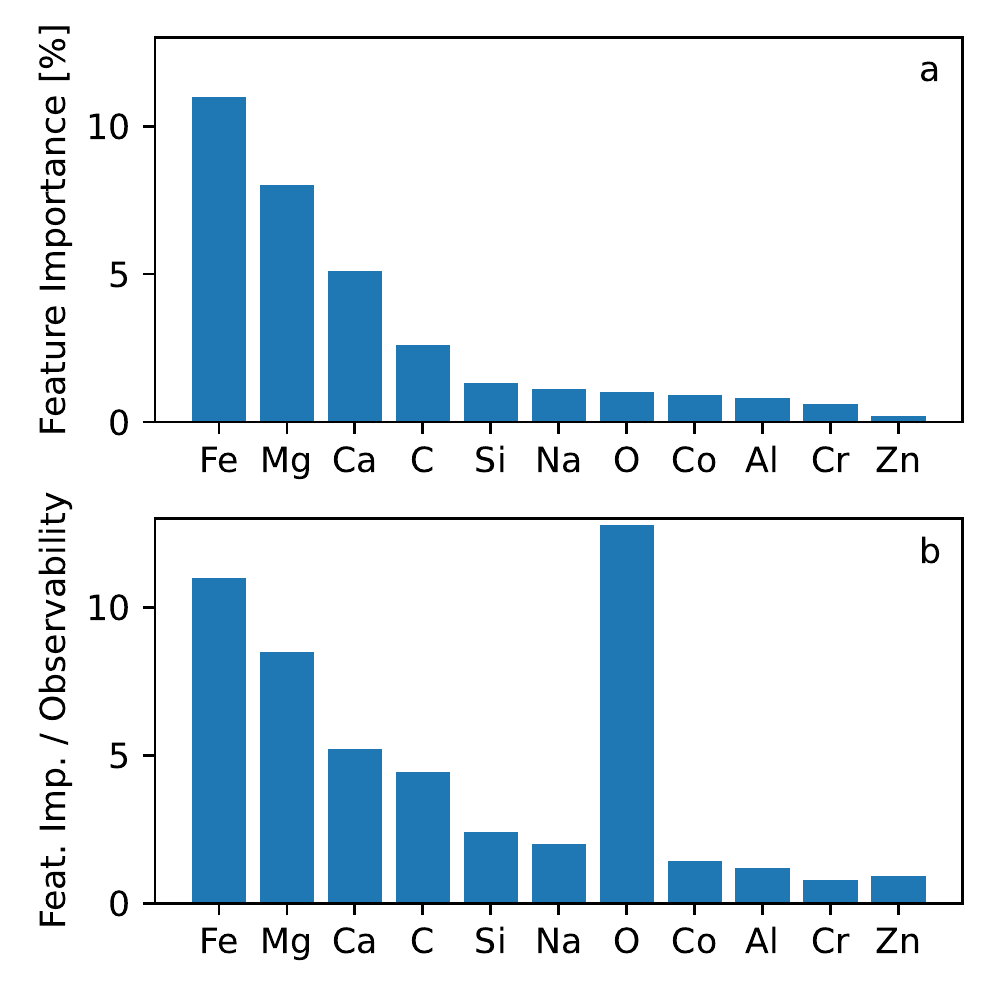}
    \caption{Panel (a): permutation feature importance for used elements. Panel (b): the feature importance divided by observability demonstrates that oxygen is informative, despite its low observability.}
    \label{fig:FI}
\end{figure}

Finally, to understand what elements are most informative in the decision process, we calculate the permutation feature importance \citep{breiman01} of the classification pipeline. The feature importance for element X is defined as the decrease in cross-validation accuracy if we randomly shuffle the values of all abundances that include element X in the cross-validation data. This score indicates how much the model depends on a specific element.
In our case, the maximum cross-validation accuracy is 70\%, and the accuracy for random guessing is 50\%. So possible values for the feature importance are in the range 0-20\%. We show the feature importance for the 11 used elements in Fig.~\ref{fig:FI}, where
the most informative elements are Fe, Mg, Ca, and C. These are the elements that are most valuable to discriminate mono- from multi-enriched EMP stars in our analysis and are usually included in observational surveys.

Two main effects dominate the feature importance of an abundance ratio: first, if an abundance ratio spans a large range in [X/Y], finding a meaningful decision boundary is easier. The training data contains scatter to mimic theoretical uncertainty. If the range over which abundance ratios in the training set are distributed is of the same order as the scatter, it is difficult to draw a reasonable decision boundary through this data. However, if the abundance ratios span a range that is significantly larger than the uncertainties, the SVM has more flexibility to identify an informative decision boundary to divide the data.
The second effect is the observability. If an element is available for most EMP stars, the SVMs likely rely on it for the classification. To differentiate the feature importance from the observability, we divided the feature importance of an element by its respective observability.
Although the order of most informative elements barely changes, we identify oxygen as very informative, relative to its low observability. Therefore, future observations of oxygen will be useful to distinguish between mono- and multi-enriched EMP stars.

\section{Discussion}
Our findings strongly indicate that a significant number of Pop~III-forming minihalos experience multiple SNe prior to EMP star formation, which suggests that most first stars formed in small clusters that contained multiple massive stars \citep{peebles68} rather than as an isolated massive star \citep{doroshkevich67}. However, it is not easy to estimate the exact multiplicity of the first stars because our result was obtained under the prior assumption that mono- and multi-enrichment are equally likely. Also, based on EMP star observations, the nature of the first stars can be investigated only for those exploded as SN \citep{ishigaki18}.

Our finding of the need for multiplicity is consistent with recent hydrodynamical simulations of Pop~III star formation, which show fragmentation of the primordial gas cloud and predict that the first stars could form in small clusters \citep{clark11,hirano17b}, resulting in multiple Pop~III SNe per minihalo. The number of fragments in a minihalo increases with time after the formation of the first protostar, and the number of Pop~III protostars per minihalo is expected to be 10--50 \citep{susa19}. However, no numerical approach has simulated the formation process until the main sequence stage of Pop~III stars and, hence, it was not possible to draw conclusions regarding the final masses and multiplicity of Pop~III stars.

Our result that Pop~III stars form in clusters is also supported by observations in the present-day Universe. At Solar metallicity, we see that the binary frequency increases with stellar mass \citep{lada06,duchene13,jason13} and that most massive stars form in binaries or higher-order systems systems \citep{zinnecker07,lee20}. Moreover, the close binary fraction seems to be anti-correlated with metallicity \citep{moe19}. Since we also expect metal-free stars to be massive, we can therefore expect that also they form in binaries, which requires more than one Pop~III star per minihalo.

\subsection{Understanding the Decision Process}
\begin{figure}
	\includegraphics[width=\columnwidth]{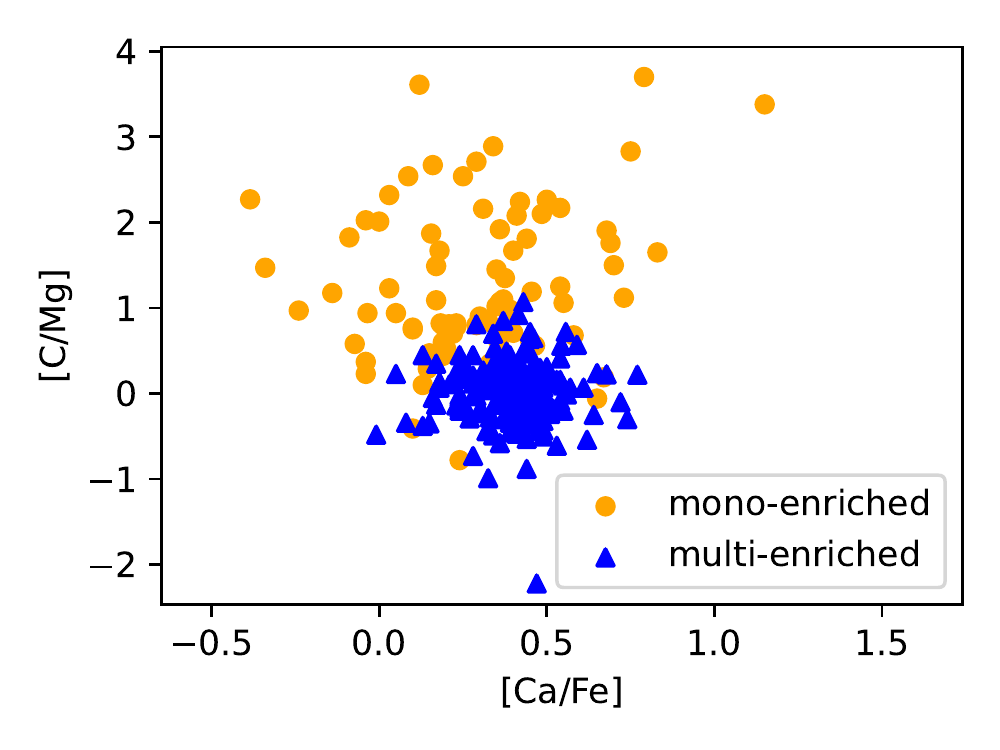}
	\includegraphics[width=\columnwidth]{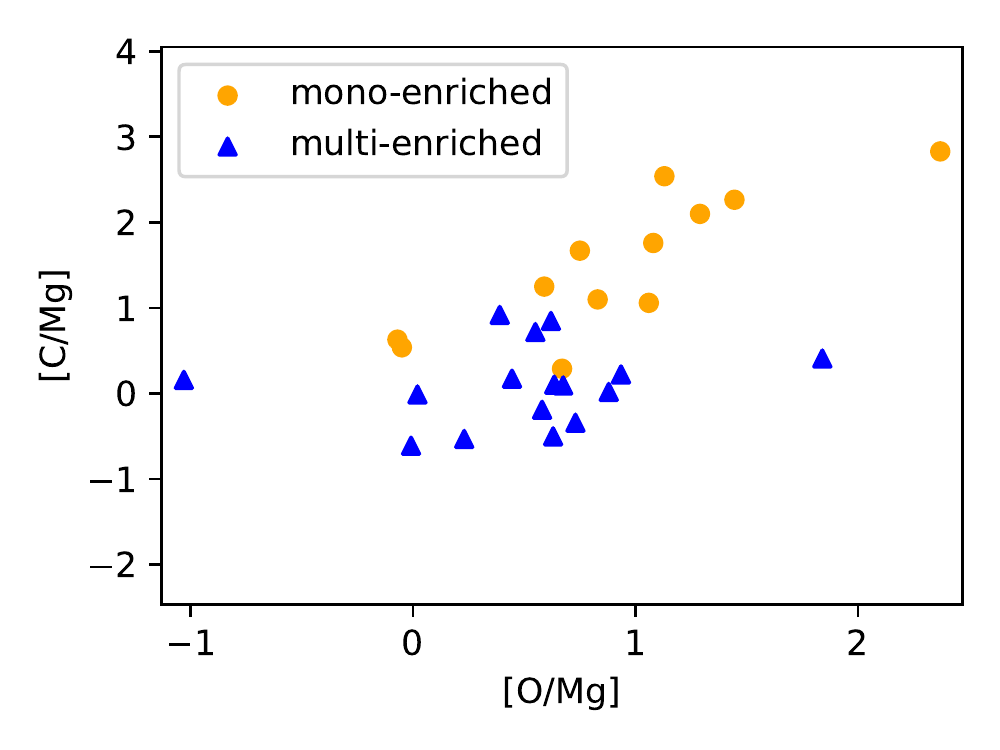}
    \caption{Illustration of the final classification in different 2D projections. The top panel uses the four most informative elements and the right panel uses oxygen, which is more challenging to observe for EMP stars. These are 2D projections of a higher-dimensional Ensemble Learning classification. Therefore, the decision boundary does not appear as smooth line in these representations.}
    \label{fig:2Dvis}
\end{figure}

We use the abundance ratios [C/Mg] and [Ca/Fe] to illustrate the final classification in Fig.~\ref{fig:2Dvis}, because these dimensions provide a high permutation importance.
Generally, mono-enriched stars are located in the outskirts of the sampled region and multi-enriched stars are more centrally concentrated. This trend is expected because multi-enrichment results in a weighted average of abundance ratios. Therefore, only mono-enriched stars can be found at extreme abundance ratios.

This figure also illustrates that two abundance ratios are not sufficient to confidently classify EMP stars. In these 2D projections, there are degenerate regions in which we find both mono- and multi-enriched EMP stars, e.g., around [Ca/Fe] $\sim 0.4$ and [C/Mg] $\sim 0.5$. However, these illustrations enable us to identify general trends, such as all EMP stars at [Ca/Fe] $<0$ are mono-enriched and most EMP stars around [Ca/Fe] $\sim 0.4$ can be multi-enriched. Moreover, we recover the trend found earlier \citep{hartwig18a} that EMP stars with [C/Mg] $\gtrsim 1$ are mostly mono-enriched.

\subsection{Comparison to Previous Works}
\citet{kobayashi11} used the elemental abundance patterns of only a few DLAs to find that faint SNe seems to be main enrichment source rather than pair-instability SNe.
\citet{welsh19} have analyzed the chemical composition of the 11 most metal-poor DLA systems known at redshift $z<5$. They use a stochastic model to infer the number of SNe that have contributed to the chemical enrichment of these systems. In contrast to EMP stars, DLAs provide a more direct way to study the chemical composition of gas in the early Universe \citep{zou20}. \citet{welsh19} find that these near-pristine gas clouds are enriched by $\lesssim 72$ SNe from massive stars. While the redshift of these DLAs ($2.6 \leq z \leq 5.0$) may be too low and their metallicity ($-3.5 \leq$ [Fe/H] $\leq -2.0$) may be too high to favour enrichment by only Pop~III SNe, their analysis shows that metal-poor gas at high redshift is enriched by multiple SNe. In a similar analysis, \citet{welsh21} analyze the stochastic enrichment of metal-poor stars in the MW halo with metallicities of [Fe/H] $\leq -2.5$. This metallicity range might include enrichment from Pop~II SNe \citep{ji15a,ishigaki21}, to which their model is also sensitive. They find that these stars are enriched by $5^{+13}_{-3}$ SNe, which supports that early star formation occurs in clusters. However, because of their metallicity range, their results do not allow a clear conclusion about the Pop~III multiplicity. In our study, we therefore focus on EMP stars with metallicities of [Fe/H] $\leq -3.0$, as we motivated above.

Compared to previous studies, our method and results are new in several regards. Previous attempts at classifying mono- and multi-enriched EMP stars used only few abundance ratios \citep{hartwig18a,hartwig19b,welsh21}. In contrast, our new method is data-driven and maximizes the information gain from all observed abundances. Previous studies used a small, biased subset of metal-poor stars or included stars at [Fe/H] $>-3$ \citep{placco18,rasmussen20,hansen20,purandardas21}. Therefore, these studies are not representative of enrichment by Pop~III SNe. In summary, our method is the first data-driven analysis of a representative sample of EMP stars for which the enrichment was dominated by Pop~III SNe.

\subsection{Prior Dependence}
\label{sec:prior}
\begin{figure}
\centering
	\includegraphics[width=\columnwidth]{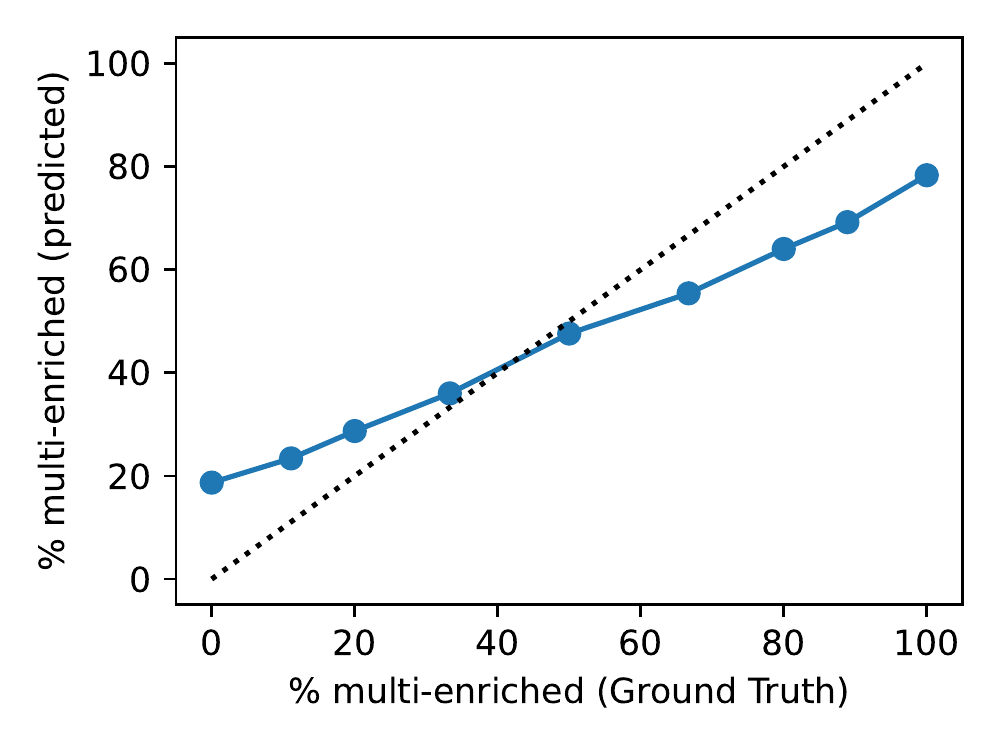}
    \caption{Predicted fraction of multi-enrichment as a function of multi-enrichment in the validation data. All these results were obtained with our fiducial model that was trained on a data set that contains equal amounts of mono- and multi-enriched stars. The black dotted line shows the diagonal to guide the eye.}
    \label{fig:MultiPred}
\end{figure}

To develop our fiducial model, we have to assume an initial distribution of mono- and multi- enriched stars. Supervised classification algorithms are most robust when trained on balanced data sets, and thus we assume an equal distribution, i.e., 50\% each, for our training set. This could affect $p_\mathrm{multi}$, and we estimate the dependence as follows. First, we use our fiducial classification pipeline for stars that are at least one standard deviation away from the decision boundary, and apply it to validation data with different fractions of mono- and multi-enriched mock observations. We then check which fraction of validation samples was classified as multi-enriched as a function of the multi-enriched fraction in the validation data. The results can be seen in Fig.~\ref{fig:MultiPred}.

Our fiducial model, which was trained under the assumption of 50\% multi-enrichment, can predict a wide range of multi-enriched fractions from $20\%-80\%$, once confronted with the data. Our model can also consistently reproduce the tendency, showing a linear trend in this figure. However, up to 20\% of multi enriched stars are missclassified. In an extreme case, even if all validation data comes from one class, about $20\%$ of samples can be missclassified. This result is related to the overall accuracy of our model of $\sim 80\%$. If all EMP stars in nature were mono-enriched, our model would still predict a multi-enriched fraction of 20\%. Instead, we find a fraction of multi-enriched EMP stars of about $70\%$, which indicates that most EMP stars are multi-enriched.

\begin{figure}
\centering
	\includegraphics[width=\columnwidth]{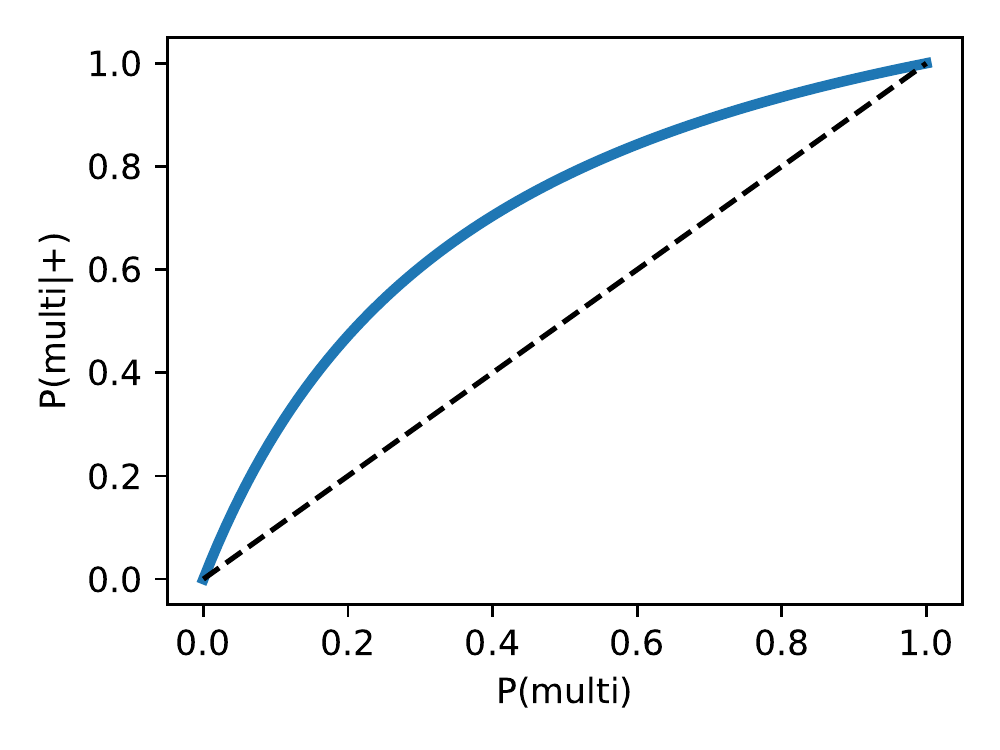}
    \caption{Probability that an EMP star that is classified as multi-enriched is actually multi-enriched as a function of the assumed multi-enriched fraction. The dashed diagonal line should guide the eye and emphasize the convex shape of the blue line.}
    \label{fig:PriorMulti}
\end{figure}

Let us now calculate how reliably we can classify an EMP star to be multi-enriched as a function of the prior assumption. The unknown fraction of multi-enriched EMP stars is $P(\mathrm{multi})$. We denote the probability that an EMP star classified as multi-enriched (+) is actually multi-enriched as $P(\mathrm{multi}|+)$. Moreover, the probability that we classify a multi-enriched star as multi enriched is $P(+|\mathrm{multi})$. The probability that we classify any star as multi-enriched is $P(+)$. Using Bayes Theorem, we can calculate the reliability of our multi-enriched predictions as
\begin{equation}
\begin{aligned}
    &P(\mathrm{multi}|+) = \frac{P(+|\mathrm{multi})P(\mathrm{multi})}{P(+)}\\
    &= \frac{P(+|\mathrm{multi})P(\mathrm{multi})}{P(+|\mathrm{multi})P(\mathrm{multi})+P(+|\mathrm{mono})P(\mathrm{mono})}
\end{aligned}
\end{equation}
All components of the right side of this equation are known from the confusion matrix (see Fig.~\ref{fig:conufs}), except for $P(\mathrm{multi})$ and $P(\mathrm{mono}) = 1- P(\mathrm{multi})$. Following  \citet{bottrell22}, we plot this function in Fig.~\ref{fig:PriorMulti} as a function of the unknown $P(\mathrm{multi})$.
It is convex and always above the diagonal. If we would relax our prior assumption of $P$(multi)$=0.5$ and allow $0.16 \leq P$(multi)$ \leq 0.84$ (central $64\%$), then the possible range of $P(\mathrm{multi}|+)$ would be $0.41-0.95$. This means that the probability for a star that is classified as multi-enriched to be actually multi-enriched is $>41\%$, even if we allow variations in the prior assumption.

While our supervised machine learning model depends on the prior assumption $P(\mathrm{multi})$, this analysis helps to understand its quantitative influence and to correct the classification results in light of better future prior assumptions. For example, if a better estimate for a prior of this classification problem is available in the future, one can use these calculation to update our results.

If we start with the fair prior assumption of $P$(multi)$=0.5$ for the ensemble and assume that for each individual star, the prior probability for multi enrichment is flat between $0-100\%$, then the marginalized distribution of $P(\mathrm{multi}|+)$ corresponds to the posterior distribution for the probability that an EMP star is multi-enriched, given that we classify it as multi-enriched. This posterior is skewed towards multi-enrichment, which supports the conclusion that the majority of EMP stars are multi-enriched.

\subsection{Variations of the Input Yields}
\label{sec:VaryYields}
We generate our training and test data based on theoretical Pop~III SN yields. There is no independent method to confirm if the distribution of SN yields is realistic. To verify if our assumptions and training data are reasonable, we confirm that the final results of our study are sufficiently robust with respect to the exact choice of Pop~III SN yields.

The first test is unphysical, but provides intuition how the prediction might change with different distributions of input yields. As we saw previously, the distributions of abundances are not identical between the observed EMP stars and the mock data that we use for training. In an attempt to equalize these distributions, we perform two transformations to minimize this discrepancy. First, we shift the distributions of mock observations so that their mean value is identical to the mean value of EMP stars. Second, in addition to the shift, we also scale the mock observations so that they have the same standard deviation as the observed abundance ratios of EMP stars. As mentioned above, this shift and scaling is not physical and should only demonstrate how robust the model is with with respect to changes in the input data.

For the second test, we increase and decrease the number of Pop~III yields that we include to generate the mock data. In the fiducial model, we select theoretical yields that have $\chi ^2 < 15.1$, where $\chi ^2$ quantifies how well these yields fit individual EMP stars \citep{ishigaki18}. We ran one more restrictive case with yields that fulfil $\chi ^2 < 13$ and one less restrictive case with $\chi ^2 < 17$. These two cases roughly halved and doubled the number of included Pop~III yields.

\begin{figure}
	\includegraphics[width=\columnwidth]{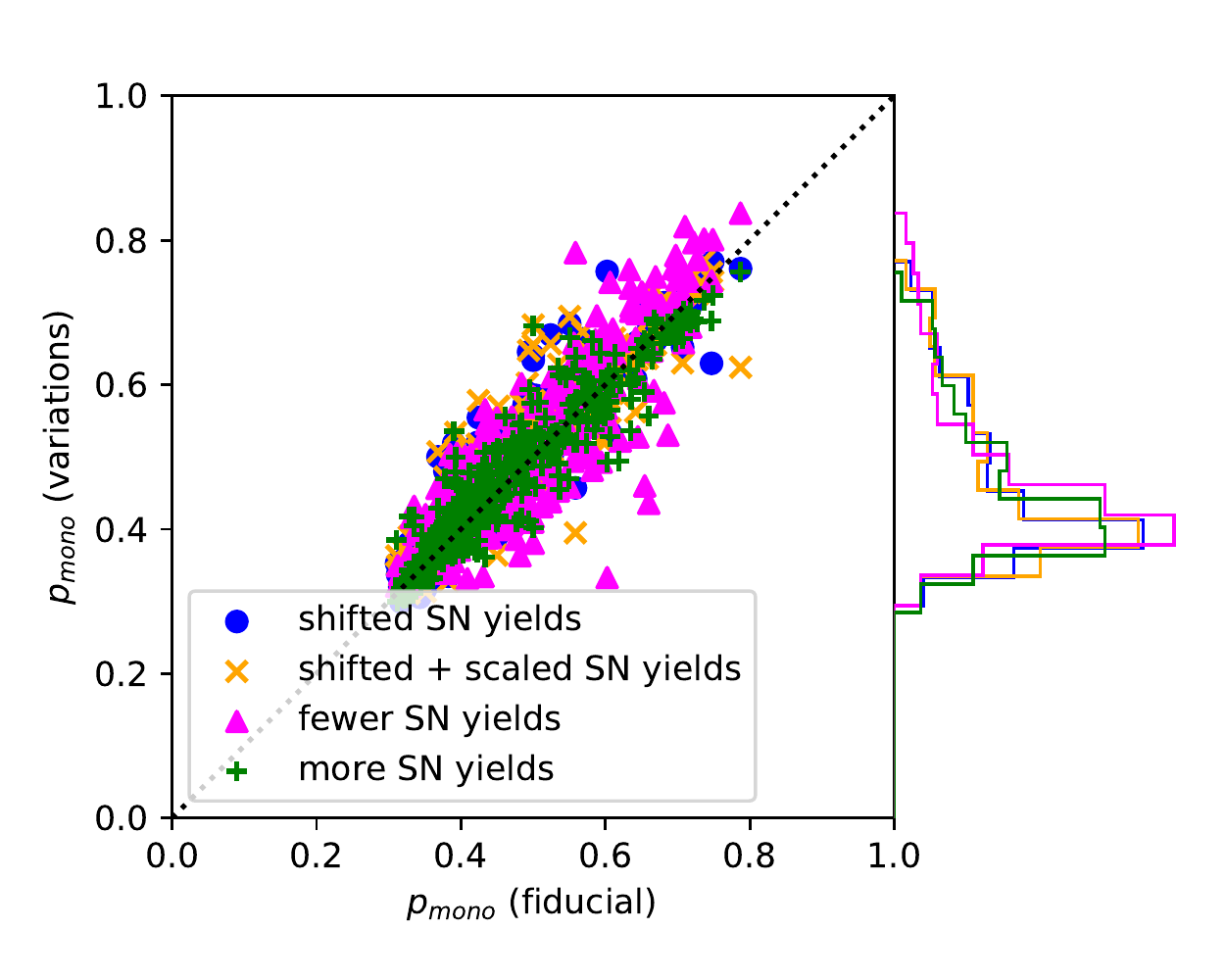}
    \caption{Predicted $p_\mathrm{mono}$ for all EMP stars. The vertical axis shows the value of four variations and the horizontal axis shows the value in the fiducial model. Most points are close to the diagonal line which illustrates that the final prediction for EMP stars is robust with respect to the exact selection of input yields. The vertical histograms on the right show the marginal distributions of the four variations.}
    \label{fig:VaryYield}
\end{figure}

The results of these tests can be seen in Fig.~\ref{fig:VaryYield}.
We compare the predicted $p_\mathrm{mono}$ for all EMP stars between the fiducial model and the two variations. In most cases, the prediction for an EMP star from the fiducial model and from one of the variations are very similar, i.e., most points are close to the diagonal. Most importantly, there are no catastrophic failures with classifications far from the diagonal.

\subsection{Enrichment with Neutron Capture Elements}
The SVMs were trained to classify EMP stars based on how many Pop~III CCSNe have enriched the gas out of which they formed. The classification is based on elemental abundances up to Zn. However, there are other channels for chemical enrichment that we did not consider explicitly and that may produce elements heavier than Zn. As two representative abundance ratios, we analyze [Ba/Fe] and [Eu/Fe] as typical tracers for the s-process and r-process, respectively. While our model does not use these abundance ratios for the classification, this information is available for some observed EMP stars, which allows to compare our predictions to the abundances of neutron-capture elements.

If these alternative enrichment channels, which we do not explicitly account for, do not provide significant amounts of elements lighter than Zn, our approach is robust (but see \citet{yong21} for the co-production). Because then we account for all enrichment channels (specifically Pop~III CCSNe) that should dominate the chemical composition of EMP stars. However, we need to be careful regarding binary mass transfer from a companion star \citep{suda04,arentsen19}. This mass transfer can add s-process elements (such as Ba), but also carbon to the EMP stars of interest. For an EMP star that was enriched in carbon via binary mass transfer, our training set, which includes carbon only from Pop~III CCSNe, is not representative anymore.

One could try to exclude EMP stars from the analysis that are highly enriched in neutron capture elements or that are in close binaries. However, there is no physically motivated boundary to define such a cut and it is certainly not one threshold value of, e.g., [Ba/Fe] below which only Pop~III SNe contribute to the chemical enrichment of EMP stars. Moreover, excluding s- or r-process enriched EMP stars from this analysis could bias the result if their enrichment is dominated by Pop~III SNe. Therefore, we decided to keep EMP stars in our sample that are enriched in neutron capture elements.

\begin{figure}
	\includegraphics[width=\columnwidth]{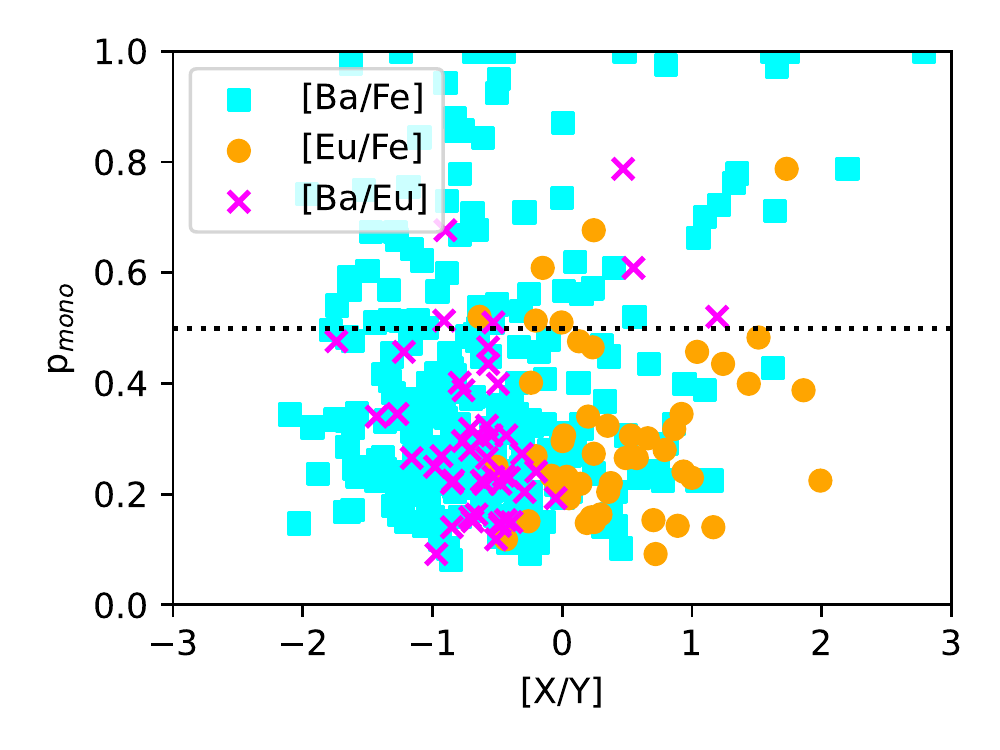}
    \caption{$p_\mathrm{mono}$ as a function of [Ba/Fe] (cyan squares), [Eu/Fe] (orange circles), and [Ba/Eu] (magenta crosses).}
    \label{fig:BaFe}
\end{figure}

In Fig.~\ref{fig:BaFe}, we show $p_\mathrm{mono}$ as a function of [Ba/Fe], [Eu/Fe], and [Ba/Eu].
Most stars for which [Eu/Fe] is available (orange points) are multi-enriched, but there is one CEMP-r star ([Eu/Fe] $\gtrsim 1$, \citealt{beers05}), which could be enriched by a single enrichment source such as magneto-rotational hypernovae \citep{yong21}.

A larger variation of $p_\mathrm{mono}$ is seen for the stars for which [Ba/Fe] is available (cyan squares); out of 12 s-process enriched EMP stars with [Ba/Fe] $>1$, only one is multi-enriched. For the stars with both Ba and Eu abundance measurements (magenta crosses in Fig.~\ref{fig:BaFe}), three stars show [Ba/Eu]$>0.5$, which suggests the s-process origin of neutron-capture elements in these stars \citep[e.g.,][]{arlandini99}. All s-process enriched EMP stars with [Ba/Fe] $>0.5$ and $p_\mathrm{mono} > 0.5$ are also carbon-enhanced with [C/Fe] $>0.7$, i.e., CEMP-s stars. Based on our results, these stars are likely to be enriched by one Pop~III CCSN. However, they might also have received heavy elements via binary mass transfer, which might attenuate their nomenclature as strictly mono-enriched.

Fig.~\ref{fig:BaFe} also shows that the majority of stars with available [Ba/Eu] are compatible with the r-process origin of neutron-capture elements in these stars \citep[e.g.,][]{arlandini99}. The origin of r-process elements is debated with various proposed enrichment channels \citep{metzger08,tanaka14,haynes19,ji19,brauer20,tarumi21,matsuno21}. In our data, we do not see any trend of $p_\mathrm{multi}$ with [Eu/Fe], which implies that our classification is agnostic with respect to the Eu abundance, which could mean that the dominant channel for the production of Eu does not produce significant amounts of elements between C and Zn.

Barium and carbon might have a similar origin and many s-process enriched stars are also C-enriched. If other enrichment channels (such as binary mass transfer) contribute to the enrichment, our SVM, which was only trained on CCSNe, might not be able to classify such unfamiliar abundances correctly. Such unfamiliar yields are rather classified as mono-enriched. Moreover, most Ba-enhanced stars are also CEMP stars, which makes them more likely to be mono-enriched (see Fig.~\ref{fig:CFe}). Therefore, s-enriched stars might appear as mono-enriched.

In summary, EMP stars that are enriched in neutron capture elements might bias our classification.  However, of the 12 stars with [Ba/Fe] $>1.0$, only one is classified as multi-enriched. So if we would exclude these stars from the classification, the fraction of multi-enriched stars would increase, which strengthens our final conclusion that most EMP stars are multi-enriched.

\section{Conclusions}
We have used supervised machine learning trained on a set of nucleosynthesis yields from \citet{ishigaki18} to classify a representative set of 462 EMP stars from the literature according to the number of SNe that have enriched the gas out of which they formed. Under the prior assumption that mono- and multi-enrichment are equally likely, we find that $31.8\% \pm 2.3\%$ of EMP stars are classified as mono-enriched. Our study is the first attempt for constraining the number of enriching SNe for EMP stars. Throughout the training, validation, and blind test process, we followed best practices for supervised machine learning and verified that the distribution of our mock observations are robust.

Our model develops a physical intuition without being explicitly trained to do so, such as the dependence of multiplicity on metallicity (Fig.~\ref{fig:NSNFeH}) and carbon enhancement (Fig.~\ref{fig:CFe}). Specifically, we find the the fraction of mono-enrichment increases from about $30\%$ at [Fe/H] $\sim -3$ to $100\%$ at [Fe/H] $\lesssim -5$. Moreover, our model offers physical explanation on the origin of the carbon abundance bimodality in EMP stars \citep{bonifacio15,yoon16}; we find that most CEMP stars at [Fe/H] $\leq -3$ are mono-enriched, which is consistent with the theory that these form out of gas that was enriched by a faint SN, and all EMP stars with [C/Fe] $>1.5$ are mono-enriched.

It may be possible to derive the exact number of SNe with a larger number of stars and elements (with errorbars) in ongoing and future spectroscopic surveys, which is the most informative observational approach to unveil the episode of the first star formation in the early Universe. We have also identified Fe, Mg, Ca, C, and O as very informative elements for this classification. Future observations of such elements and smaller uncertainties of the measured abundances will improve our predictions in the future. By training our classifier on independent SN yields from other models \citep[e.g.][]{heger10, limongi12}, we can also test if our results are independent of the assumed yields.

\begin{acknowledgments}
We thank Ralf Klessen, Louise Welsh, Naoki Yoshida, Mattis Magg, and the anonymous referee for discussions and valuable feedback on the paper draft. We are also grateful to Vinicius Placco for providing the tabulated carbon corrections. This work was supported by World Premier International Research Center Initiative (WPI), MEXT, Japan, and JSPS KAKENHI Grant Numbers JP17K05382, JP20K04024, 17K14249, 19K23437, 20K14464, and JP21H04499. C.K. acknowledges funding from the UK Science and Technology Facility Council (STFC) through grant ST/ R000905/1.
\end{acknowledgments}

%



\software{The source code was written in Python, is based on Scikit Learn, and is available online\footnote{\url{{https://gitlab.com/thartwig/emu-c}}}. Furthermore, we used python \citep{python09}, numpy \citep{harris20}, scipy \citep{virtanen20}, matplotlib \citep{hunter07}, and scikit-learn \citep{scikit}.}



\appendix

\section{Observational and Theoretical Uncertainties}
\label{sec:uncert}
Observational errors that we could obtain from observational papers often do not include the major error source, which is the NLTE effect \citep{andrievsky07, lind11, zhao16, mashonkina17, nordlander17}, and the prediction from stellar atmosphere modelling is confirmed by a galactic chemical evolution (GCE) model for Na and Al (also for K; not confirmed for Mn and Cu). The GCE model also predicts a strong NLTE effect for Cr I observations \citep{kobayashi06,sneden16}.
In addition, the 3D effect should also be included in particular for C, N, and O abundances \citep{amarsi19}, and possibly for Mg \citep{bergemann17} and Mn \citep{bergemann19}.
These effects depend on the mass, metallicities, and evolutionary stages of the observed stars.
The corrections for different elements may be correlated, but we do not have a good understanding of the correlation of the corrections among various elements.
Therefore, we assume a single matrix for all of our sample (mostly taken from the SAGA database) including all of these possible effects. The errors of abundance ratios relative to Fe are set based on previous results \citep{kobayashi20}, and we calculate the errors of the other combination of elements as the squared sum:
\begin{equation}
    \sigma_\mathrm{obs}([\mathrm{X/Y]}) = \sqrt{ \sigma_\mathrm{obs}([\mathrm{X/Fe]})^2 + \sigma_\mathrm{obs}([\mathrm{Y/Fe]})^2 }.
\end{equation}

For theoretical models, a few factors should be taken into account (see the section 3.6 of \citet{kobayashi20} for detailed discussion). The main error sources can be summarized as (1) nuclear reactions including neutrino processes, (2) stellar rotation and any mixing during hydrostatic burning, (3) mixing during SN explosion, and (4) fallback.
Apart from the effect (1), these effects are not independent, and thus the errors should not be treated as the squared sum.
As a result, there are multiple elements that are similarly affected mainly by one effect, and in that case, the errors of these elemental abundance ratios are small. Therefore, we provide a matrix of theoretical errors of all used combinations of elemental abundances in Table ~\ref{tab:uncert}. These values are estimated by comparing our stellar evolution calculations with/without mixing, and our 1D and 2D nucleosynthesis calculations with different mixing and fallback \citep{umeda00,kobayashi06,tominaga07,tominaga09,kobayashi11agb,nomoto13,kobayashi20}.

The largest error can be seen for Na and Al due to the effect (2). Among $\alpha$ elements (O, Mg, Si, and Ca), the error of [(O, Ca)/Fe] is set to be smaller than that of [(Mg, Si)/Fe]; this is suggested by the GCE model, and the reason is likely to be the effect (1).
Iron peak elements (including Ti) can be affected by all of these effects, and the impact of each effect can be evaluated at each nucleosynthetic region (a layer in 1D) inside the SN ejecta. Namely, Cr, Mn are mainly produced in the incomplete Si-burning region, while Fe, Ni, Co, and Zn are produced in the complete Si-burning region \citep{kobayashi06}, which results in the smaller errors for Cr/Mn, Ni/Fe, and Co/Zn. The errors for Co and Zn are larger due to the effects (3 and 4, possibly 2 and 1 as well) of aspherical explosions, which is included in our 1D mixing-fallback model, but may not be fully (Kobayashi \& Tominaga, in prep.).
These dependencies are confirmed by the GCE model comparing to the NLTE abundances of high-resolution observations \citep{kobayashi20}. We obtain the matrix in Tab.~\ref{tab:uncert} taken into account all of these non-linear effects in nuclear astrophysics, and the matrix can in principle be used for other nucleosynthesis yield sets.

\begin{deluxetable}{cccccccccccccc}
\tablecaption{Observational (top right triangle, [X(row)/Y(column)]) and theoretical (bottom left triangle, [X(column)/Y(row)]) uncertainties. For example, the theoretical uncertainty of [C/O] is 0.10\,dex and the observational uncertainty of [C/O] is 0.28\,dex. The observational uncertainties are based on iron. For some abundance ratios, we provide the asymmetric errors. These values should be subtracted for correcting observed/modelled values, i.e., [Na/Fe]$_{\rm obs}-^{+0.5}_{-0.2}$ and [Fe/Na]$_{\rm model}-^{+0.5}_{-0.1}$.
\label{tab:uncert}}
\tablehead{\colhead{ } & \colhead{C} & \colhead{O} & \colhead{Na} & \colhead{Mg} & \colhead{Al} & \colhead{Si} & \colhead{Ca} & \colhead{Cr} & \colhead{Mn} & \colhead{\textbf{Fe}} & \colhead{Co} & \colhead{Ni} & \colhead{Zn}}
\startdata
C   &--         &   $\pm$\,0.28    &$^{+0.28} _{-0.56}$&$^{+0.36} _{-0.28}$&$^{+0.54} _{-0.28}$&$\pm$\,0.28       &$\pm$\,0.28&$^{+0.45} _{-0.28}$&$^{+0.36} _{-0.28}$& $\pm$\,{0.20} & $\pm$\,0.28 & $\pm$\,0.28 & $\pm$\,0.28\\
O	&$\pm$\,0.10	    &	--	    &$^{+0.28} _{-0.54}$&$^{+0.36} _{-0.28}$&$^{+0.54} _{-0.28}$&$\pm$\,0.28       &$\pm$\,0.28	    &$^{+0.45} _{-0.28}$&$^{+0.36} _{-0.28}$&	$\pm$\,{0.20}	&	$\pm$\,0.28	&	$\pm$\,0.28	&	$\pm$\,0.28	\\
Na	&$^{+0.10} _{-0.50}$	&$^{+0.10} _{-0.50}$&	--	    &$^{+0.58} _{-0.28}$&$^{+0.71} _{-0.28}$&$^{+0.54} _{-0.28}$&$^{+0.54} _{-0.28}$&$^{+0.64} _{-0.28}$&$^{+0.58} _{-0.28}$&	${^{+0.50} _{-0.20}}$	&	$^{+0.54} _{-0.28}$	&	$^{+0.54} _{-0.28}$	&	$^{+0.54} _{-0.28}$	\\
Mg	&$\pm$\,0.10	&$\pm$\,0.10&$^{+0.50} _{-0.10}$&--	        &$^{+0.54} _{-0.36}$&$^{+0.28} _{-0.36}$&$^{+0.28} _{-0.36}$&$^{+0.45} _{-0.36}$&$\pm$\,0.36	    &	$^{+0.20} _{-0.30}$	&	$^{+0.28} _{-0.36}$	&	$^{+0.28} _{-0.36}$	&	$^{+0.28} _{-0.36}$	\\
Al	&$^{+0.10} _{-0.50}$	&$^{+0.10} _{-0.50}$&$\pm$\,0.20&$^{+0.10} _{-0.50}$&--	        &$^{+0.28} _{-0.54}$&$^{+0.28} _{-0.54}$&$^{+0.45} _{-0.54}$&$^{+0.36} _{-0.54}$&	$^{+0.20} _{-0.50}$	&	$^{+0.28} _{-0.54}$&	$^{+0.28} _{-0.54}$	&	$^{+0.28} _{-0.54}$	\\
Si	&$^{+0.10} _{-0.20}$	&$^{+0.10} _{-0.20}$&$^{+0.50} _{-0.20}$&$^{+0.10} _{-0.20}$&$^{+0.50} _{-0.20}$&--	        &	$\pm$\,0.28    &$^{+0.45} _{-0.28}$&$^{+0.36} _{-0.28}$&	$\pm$\,{0.20}	&	$\pm$\,0.28	&	$\pm$\,0.28	&	$\pm$\,0.28	\\
Ca	&$^{+0.10} _{-0.20}$	&$^{+0.10} _{-0.20}$&$^{+0.50} _{-0.20}$&$^{+0.10} _{-0.20}$&$^{+0.50} _{-0.20}$&$\pm$\,0.20&	--	    &$^{+0.45} _{-0.28}$&$^{+0.36} _{-0.28}$&	$\pm$\,{0.20}	&	$\pm$\,0.28	&	$\pm$\,0.28	&	$\pm$\,0.28	\\
Cr	&$^{+0.10} _{-0.15}$&$^{+0.10} _{-0.15}$&$^{+0.50} _{-0.15}$&$^{+0.10} _{-0.15}$&$^{+0.50} _{-0.15}$&$^{+0.20} _{-0.15}$&$^{+0.20} _{-0.15}$&	--	    &$^{+0.36} _{-0.45}$&	$^{+0.20} _{-0.40}$	&	$^{+0.28} _{-0.45}$	&	$^{+0.28} _{-0.45}$	&	$^{+0.28} _{-0.45}$	\\
Mn	&$^{+0.10} _{-0.20}$&$^{+0.10} _{-0.20}$&$^{+0.50} _{-0.20}$&$^{+0.10} _{-0.20}$&$^{+0.50} _{-0.20}$&$\pm$\,0.20&$\pm$\,0.20&$\pm$\,0.10       &	--	    &	$^{+0.20} _{-0.30}$	&	$^{+0.28} _{-0.36}$	&	$^{+0.28} _{-0.36}$&	$^{+0.28} _{-0.36}$	\\
\textbf{Fe}	&$\pm$\,0.10	    &$\pm$\,0.10	    &$^{+0.50} _{-0.10}$&$\pm$\,0.10&$^{+0.50} _{-0.10}$&$^{+0.20} _{-0.10}$&   $^{+0.20} _{-0.10}$&$^{+0.15} _{-0.10}$&$^{+0.20} _{-0.10}$&	--	&	$\pm$\,{0.20}	&	$\pm$\,{0.20}	&	$\pm$\,{0.20}	\\
Co	&$^{+0.10} _{-0.30}$&$^{+0.10} _{-0.30}$&$^{+0.50} _{-0.30}$&$^{+0.10} _{-0.30}$&$^{+0.50} _{-0.30}$&$^{+0.20} _{-0.30}$&$^{+0.20 } _{-0.30}$&$^{+0.15} _{-0.30}$&$^{+0.20} _{-0.30}$&$^{+0.05} _{-0.25}$&--	&	$\pm$\,0.28	&	$\pm$\,0.28	\\
Ni	&$^{+0.10} _{-0.15}$&$^{+0.10} _{-0.15}$&$^{+0.50} _{-0.15}$&$^{+0.10} _{-0.15}$&$^{+0.50} _{-0.15}$&$^{+0.20} _{-0.15}$&$^{+0.20} _{-0.15}$&$\pm$\,0.15       &$^{+0.20} _{-0.15}$&$^{+0.05} _{-0.10}$&$^{+0.25} _{-0.10}$&	--	&	$\pm$\,0.28	\\
Zn	& $^{+0.10} _{-0.30}$ & $^{+0.10} _{-0.30}$ & $^{+0.50} _{-0.30}$ & $^{+0.10} _{-0.30}$ & $^{+0.50} _{-0.30}$ & $^{+0.20} _{-0.30}$ & $^{+0.20} _{-0.30}$ & $^{+0.15} _{-0.30}$ & $^{+0.20} _{-0.30}$ & $^{+0.05} _{-0.25}$ & $\pm$\,0.20 & $^{+0.10} _{-0.25}$ &	--	\\
\enddata
\end{deluxetable}

\section{Carbon corrections}
\label{sec:Ccorr}
\begin{figure}[htb]
	\includegraphics[width=0.5\columnwidth]{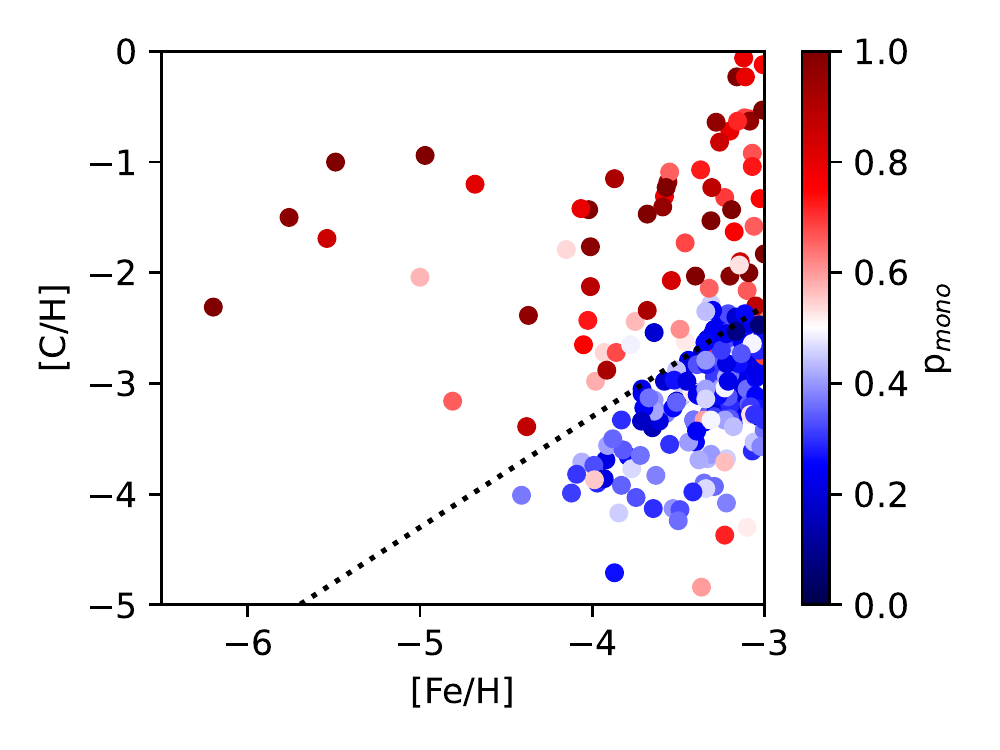}
    \caption{Same as Fig.~\ref{fig:CFe} but without carbon corrections. This means that we did not use the carbon corrections for the classification, nor for the vertical axis.}
    \label{fig:CEMP_notCorr}
\end{figure}

While the carbon corrections do not affect the mean ensemble classification, they slightly affect the distribution in the [C/H] vs. [Fe/H] plane. To allow readers a transparent comparison, we also present the results without carbon correction in Fig.~\ref{fig:CEMP_notCorr}.
In this representation, the fiducial CEMP boundary at [C/Fe] $=0.7$ (dotted line) seems to better discriminate between mono- and multi-enrichment.

\section{Decision Maps}
To make our results available for the community, we made the source code public\footnote{\url{{https://gitlab.com/thartwig/emu-c}}}. For a direct, qualitative comparison to our model, we provide additional classification maps in this section. These maps show 2D projections of the final classification (same structure as Fig.~\ref{fig:2Dvis}).
\label{sec:DecMaps}
\begin{figure*}[htb]
	\includegraphics[width=0.49\columnwidth]{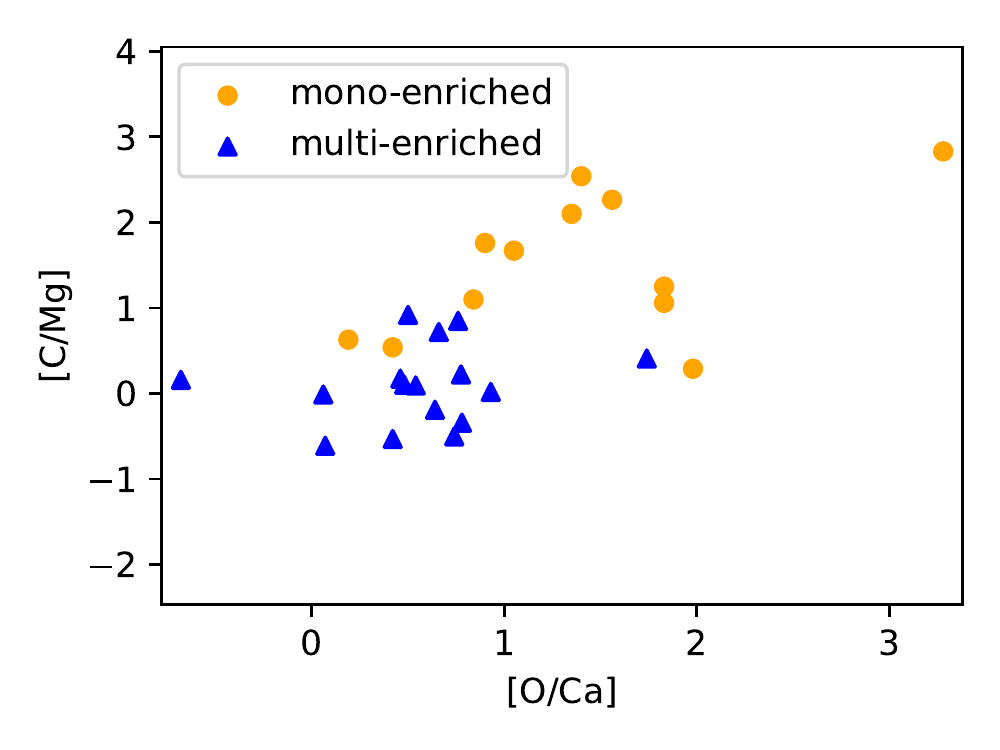}
	\includegraphics[width=0.49\columnwidth]{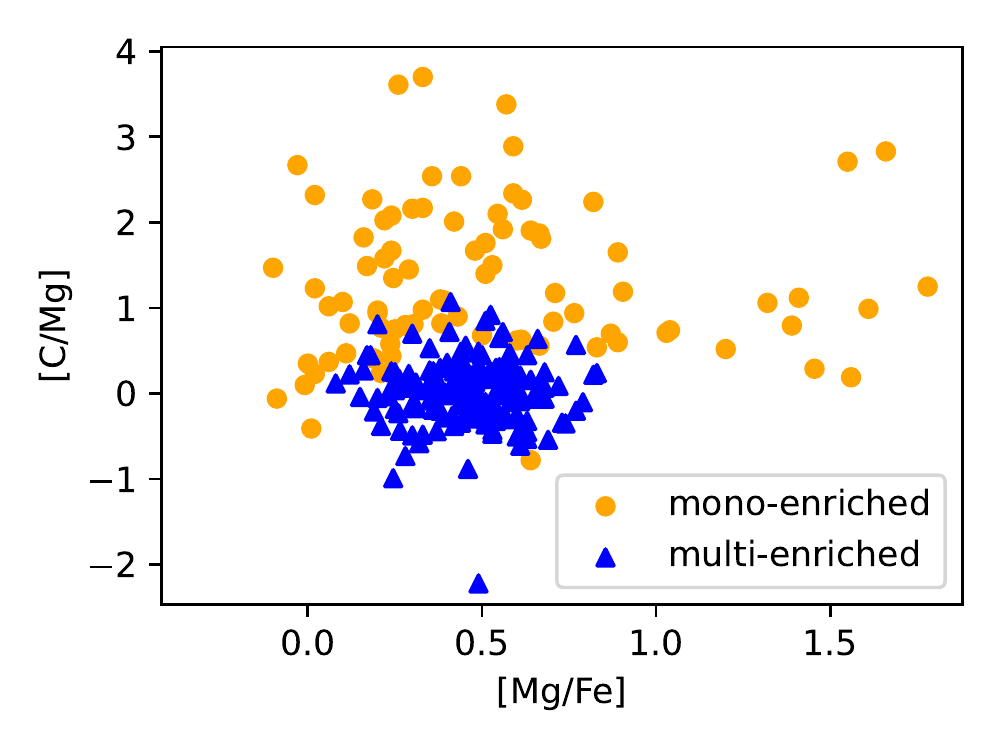}
	\includegraphics[width=0.49\columnwidth]{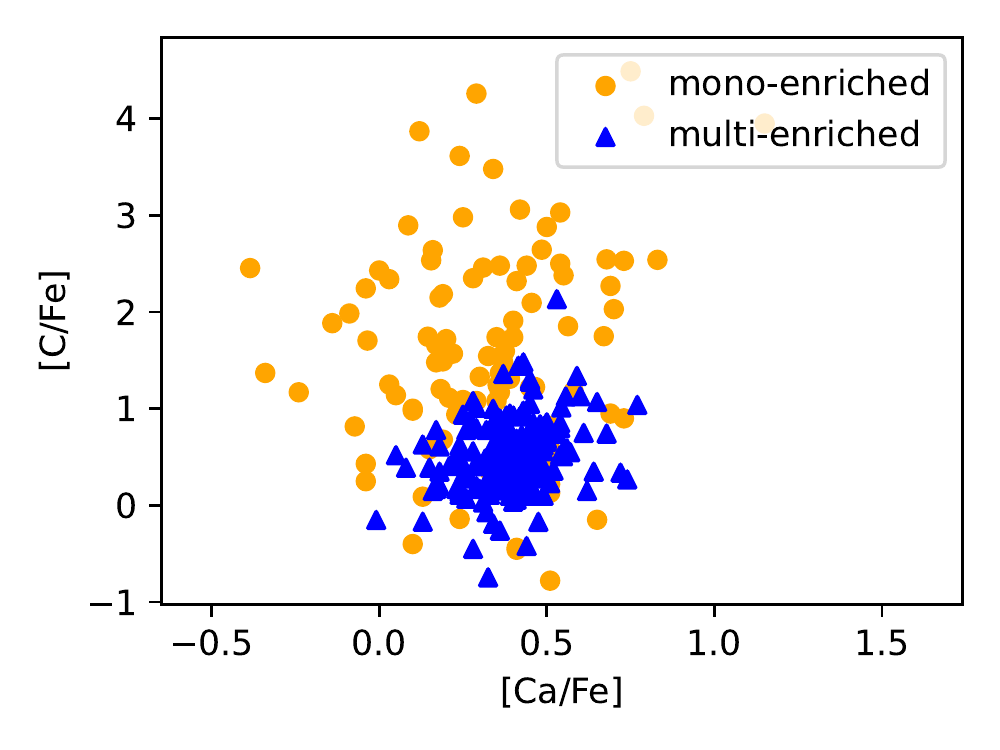}
	\includegraphics[width=0.49\columnwidth]{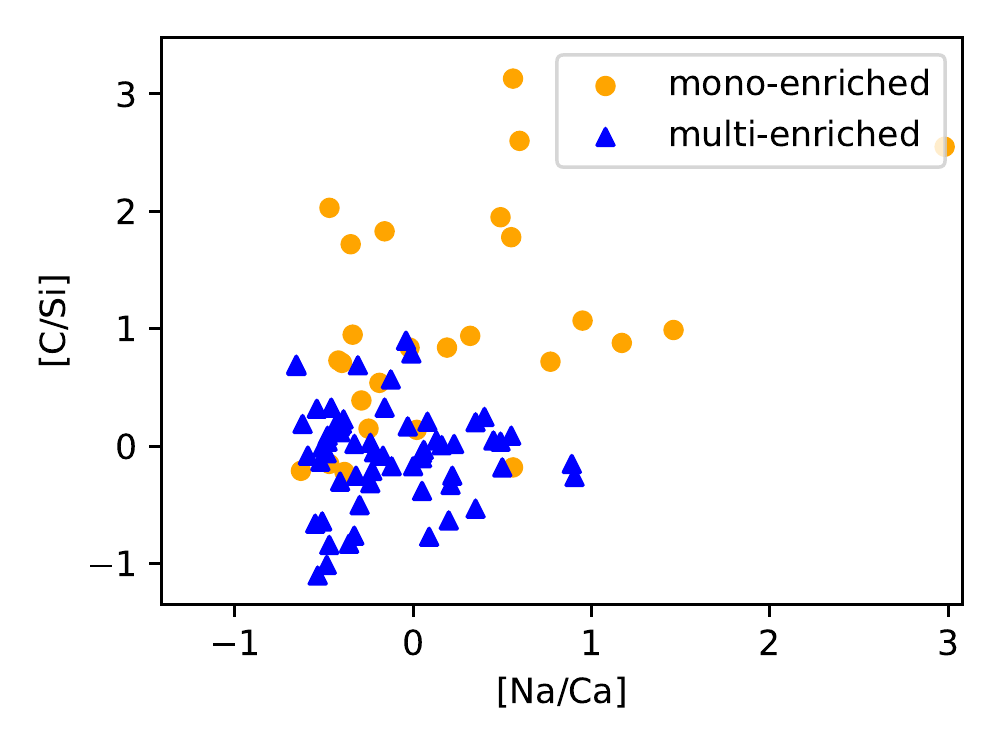}
    \caption{2D classification maps for all EMP stars for which these abundances are available.}
    \label{fig:2DvisApp}
\end{figure*}


\section{Mock Histograms}
In addition to the three distributions in Fig.~\ref{fig:HistTrain}, we provide here the remaining histograms of abundances that are used in the classification.
\label{sec:MockHist}
\begin{figure}
	\includegraphics[width=0.29\textwidth]{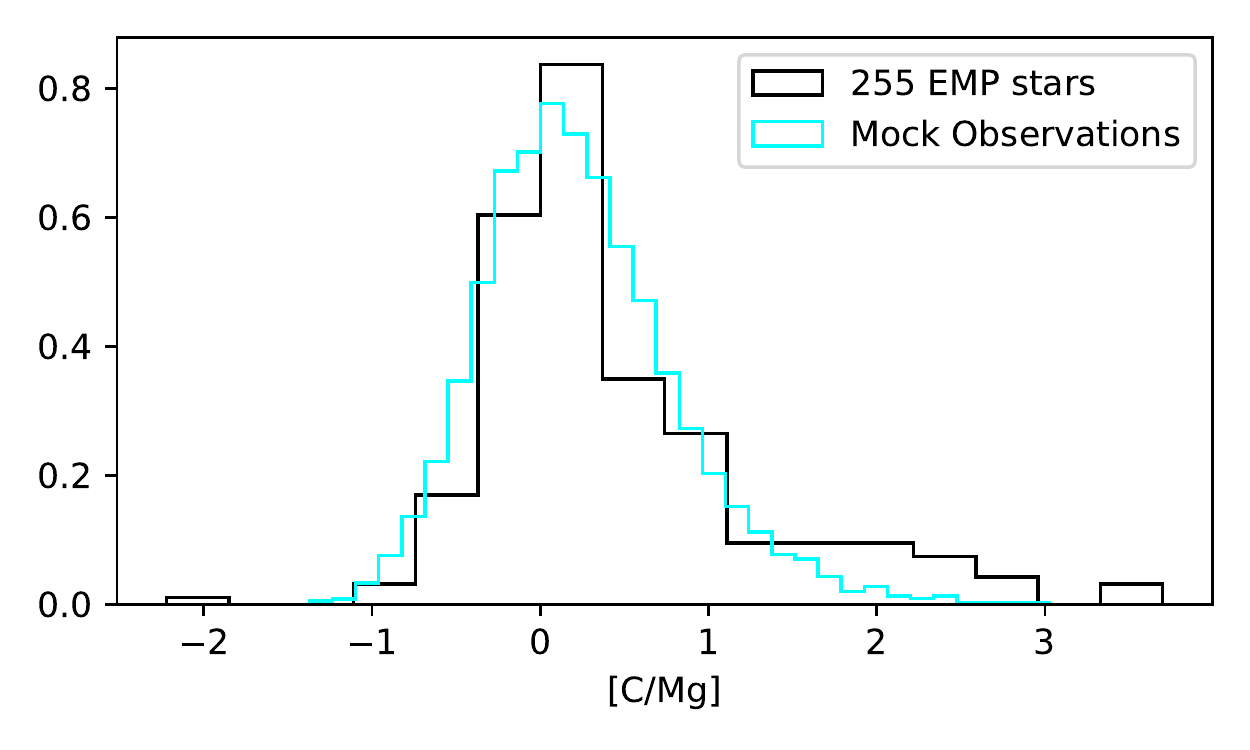}
	\includegraphics[width=0.29\textwidth]{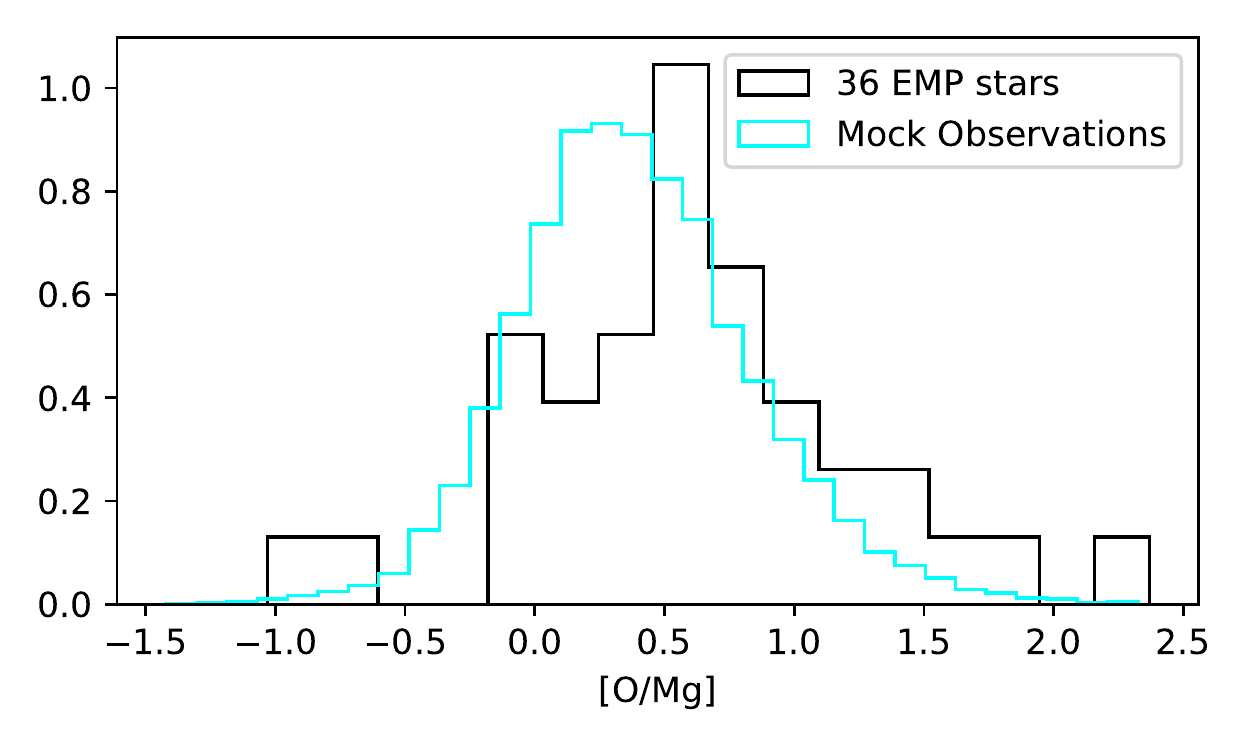}
	\includegraphics[width=0.29\textwidth]{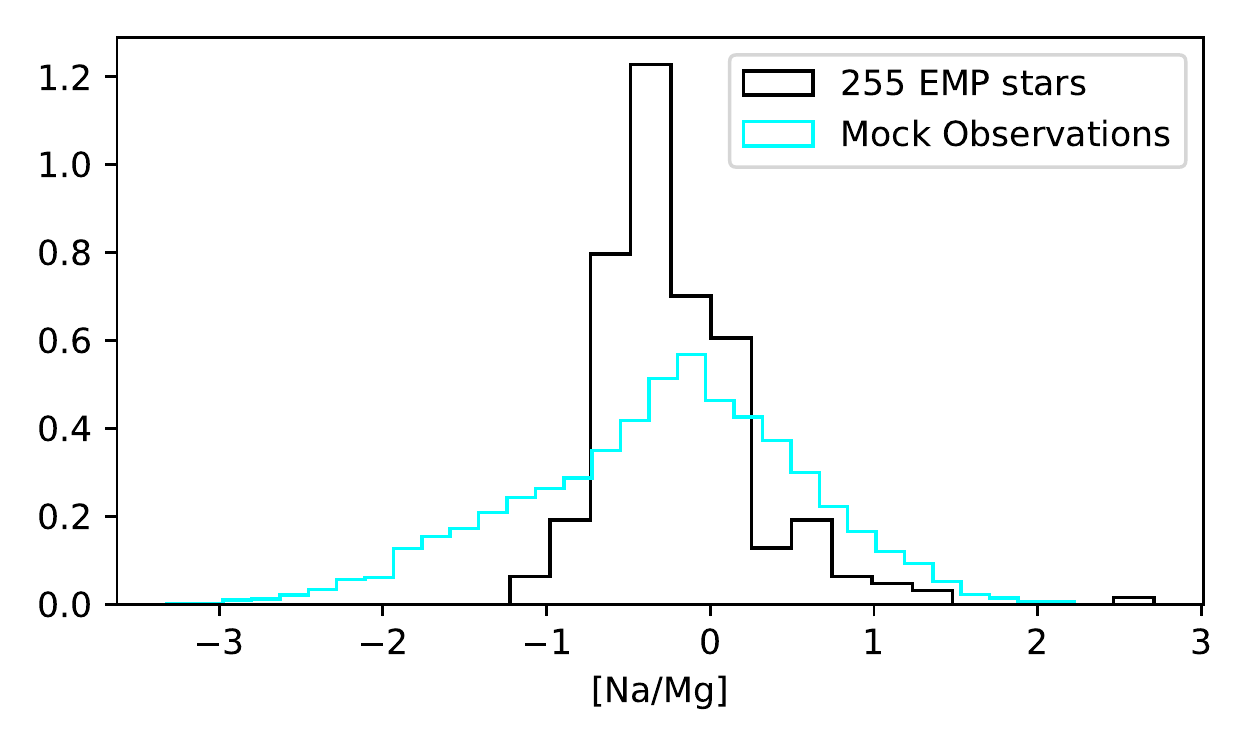}
	\includegraphics[width=0.29\textwidth]{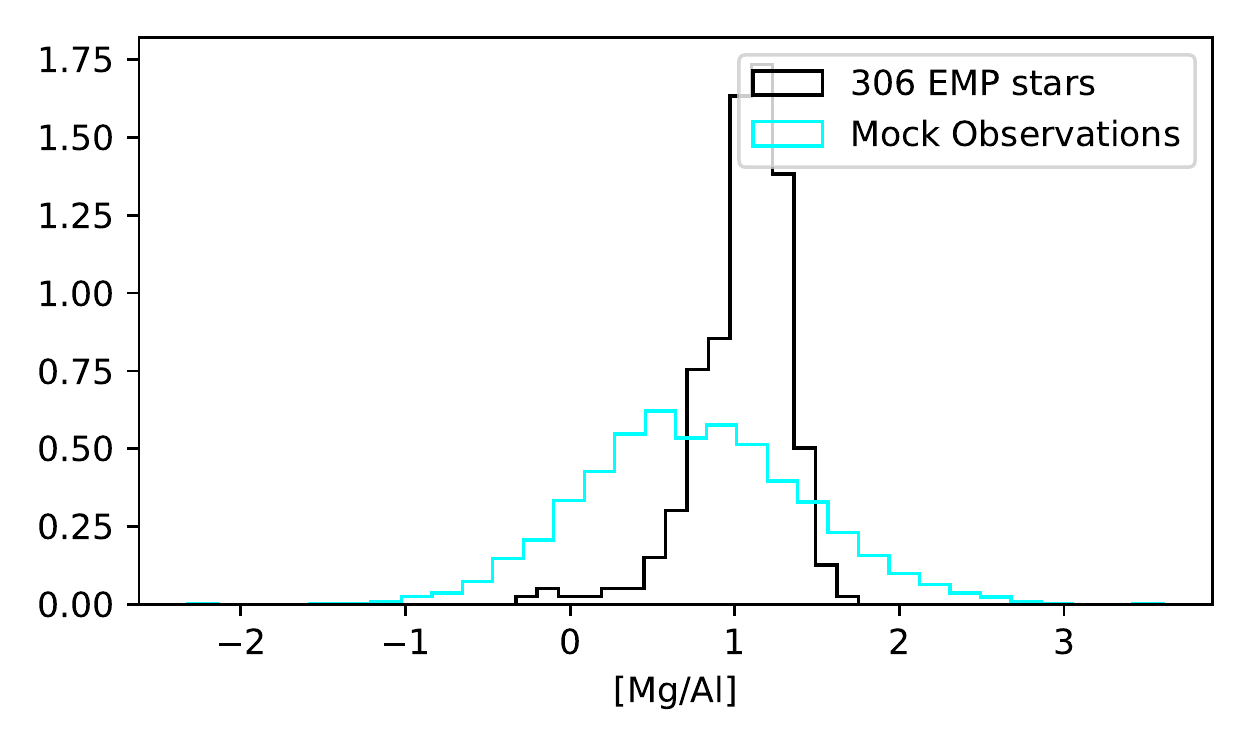}
	\includegraphics[width=0.29\textwidth]{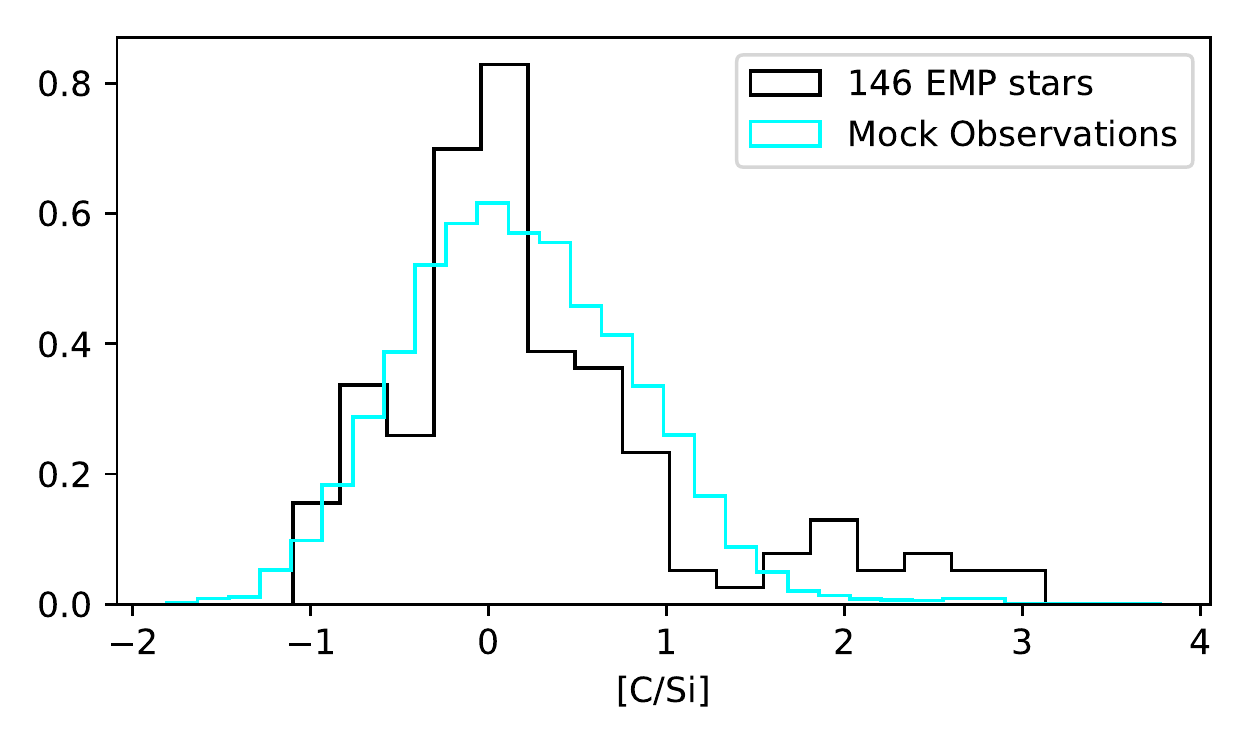}
	\includegraphics[width=0.29\textwidth]{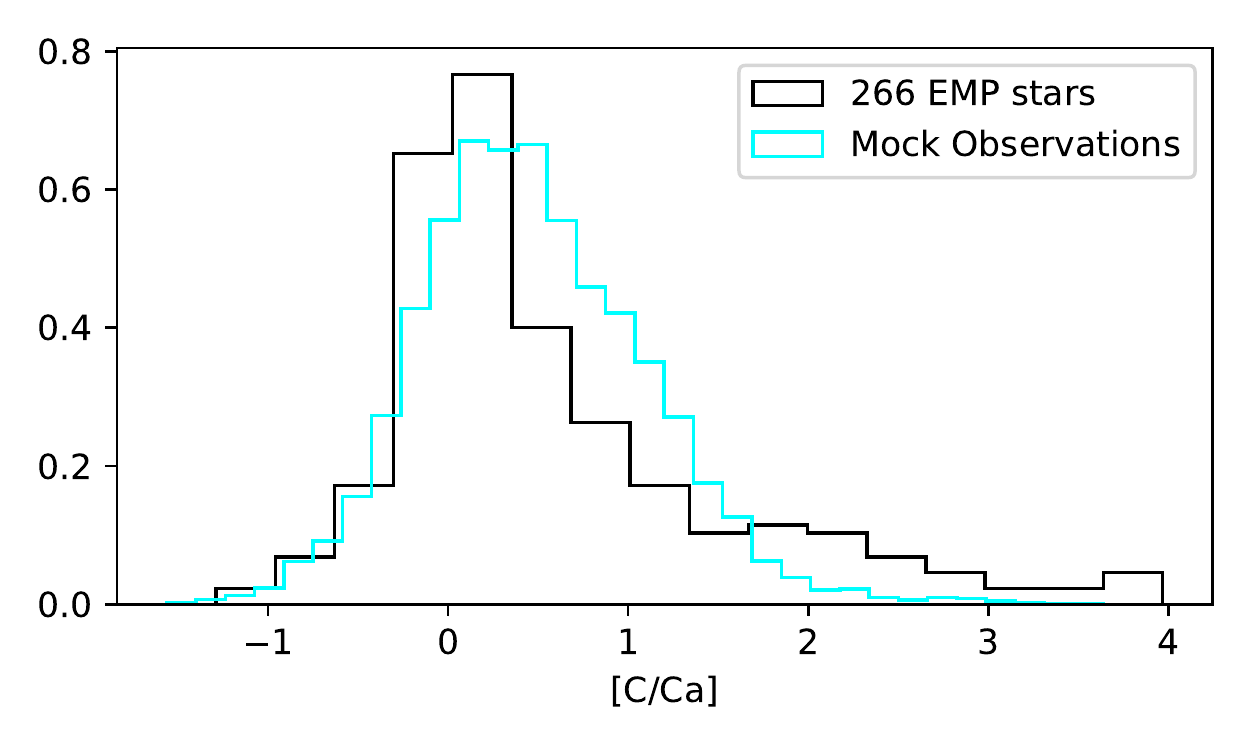}
	\includegraphics[width=0.29\textwidth]{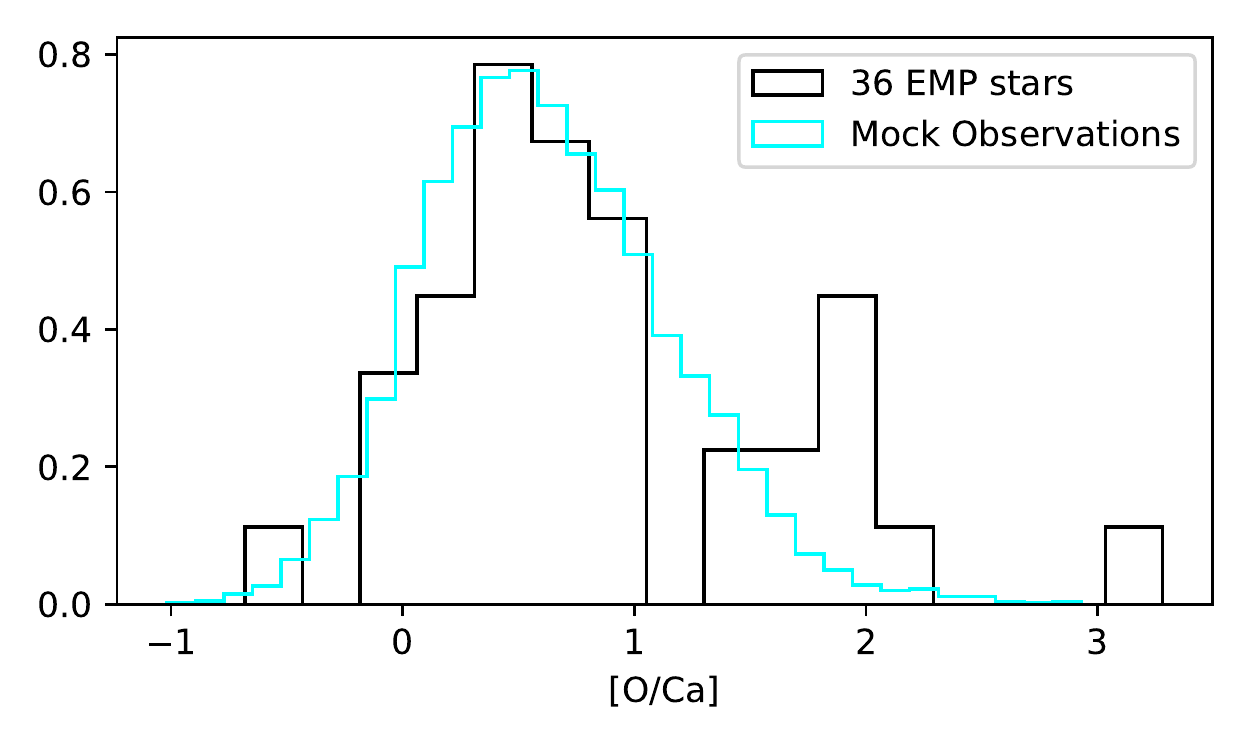}
	\includegraphics[width=0.29\textwidth]{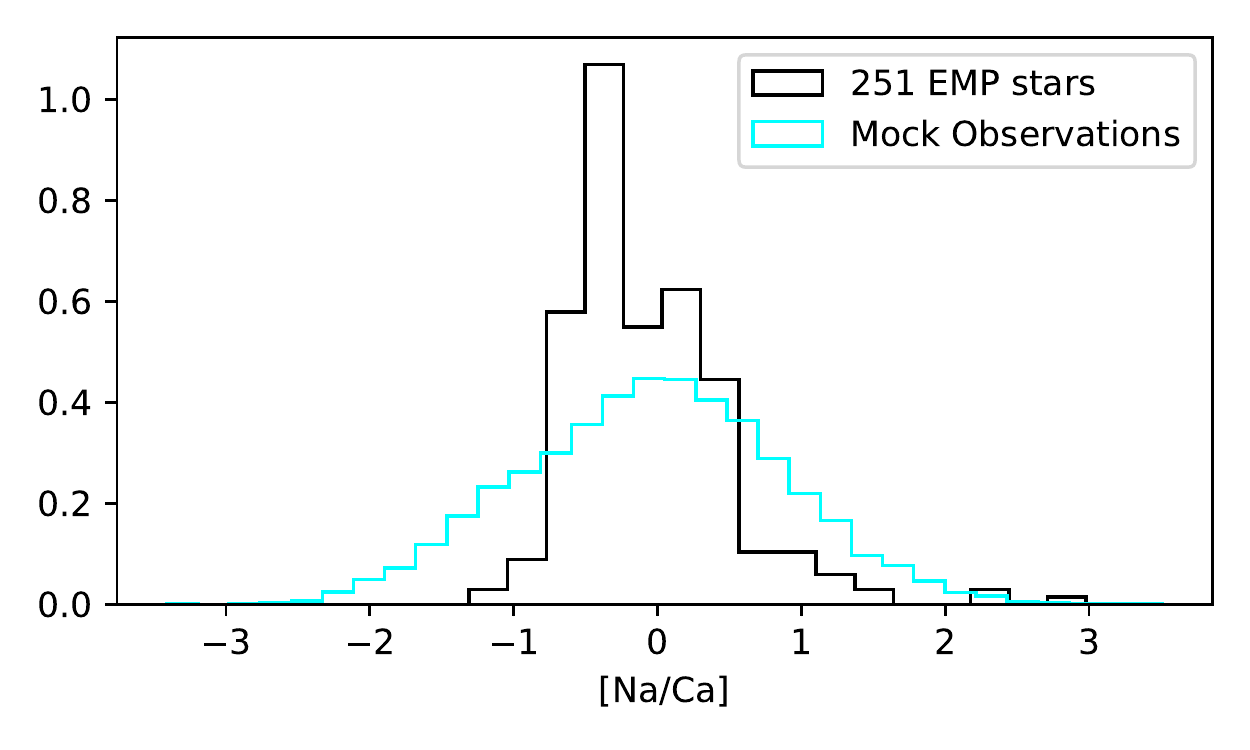}
	\includegraphics[width=0.29\textwidth]{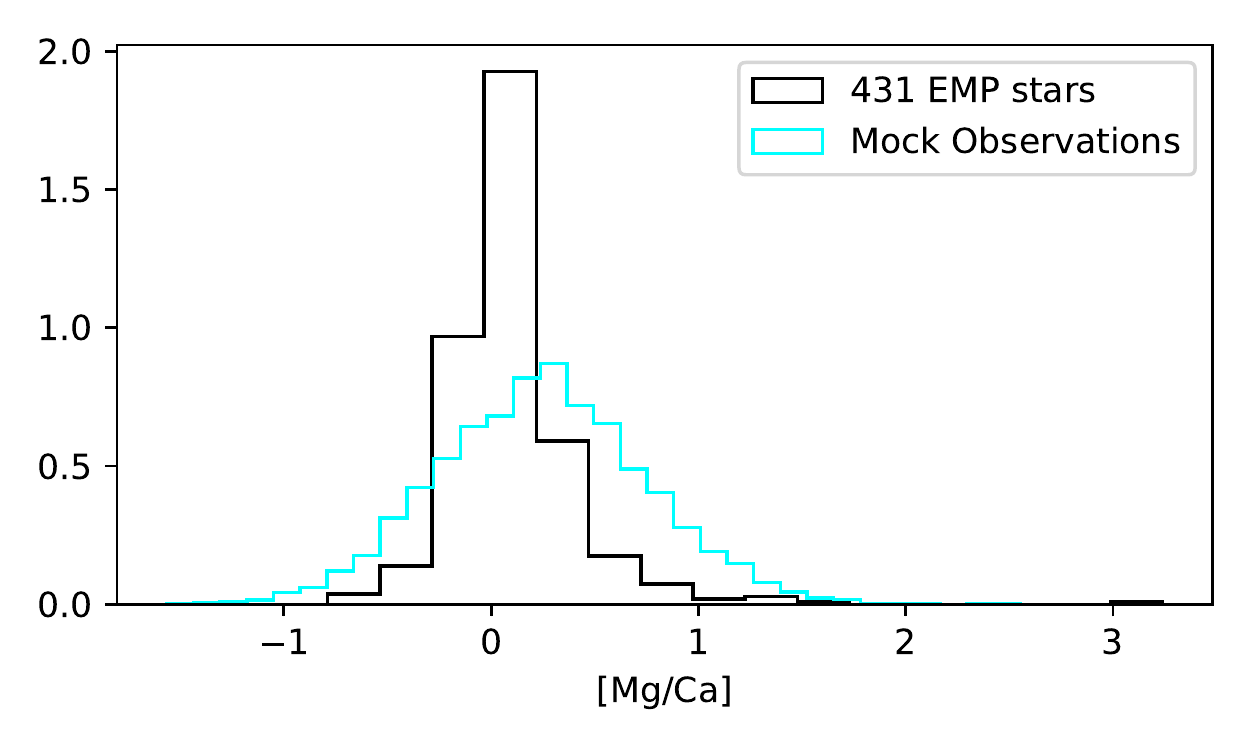}
	\includegraphics[width=0.29\textwidth]{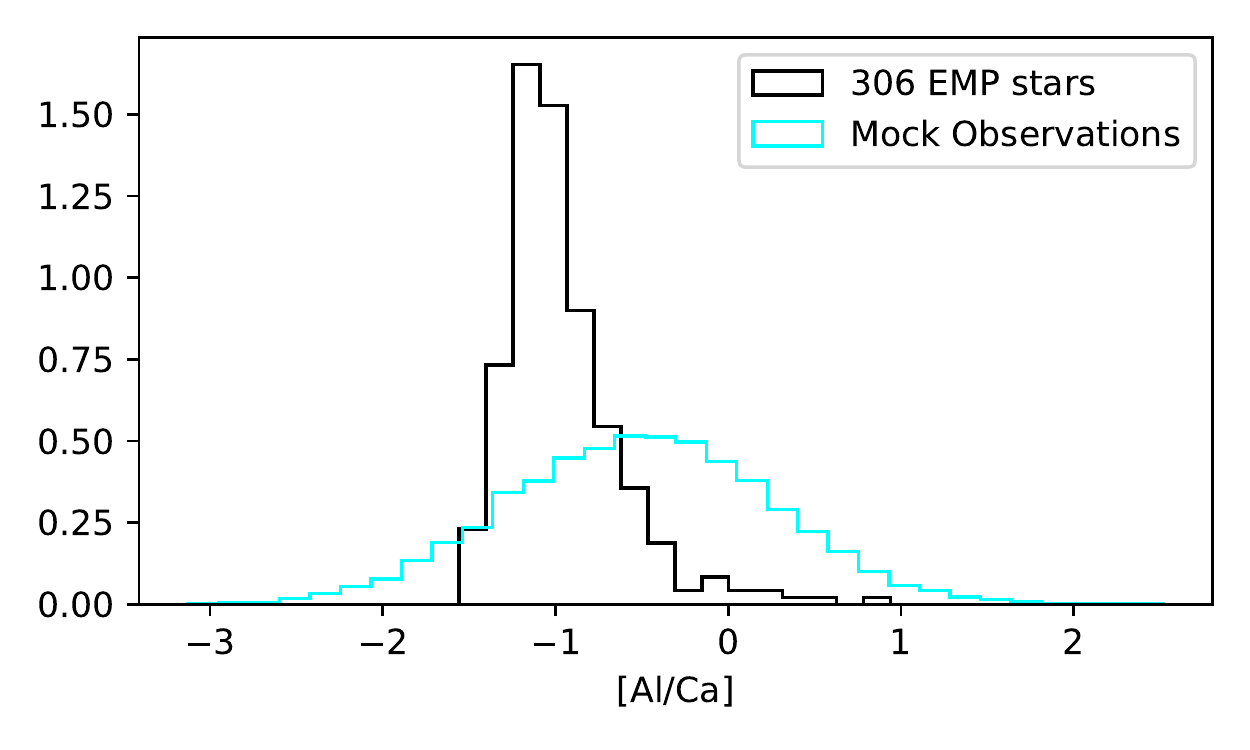}
	\includegraphics[width=0.29\textwidth]{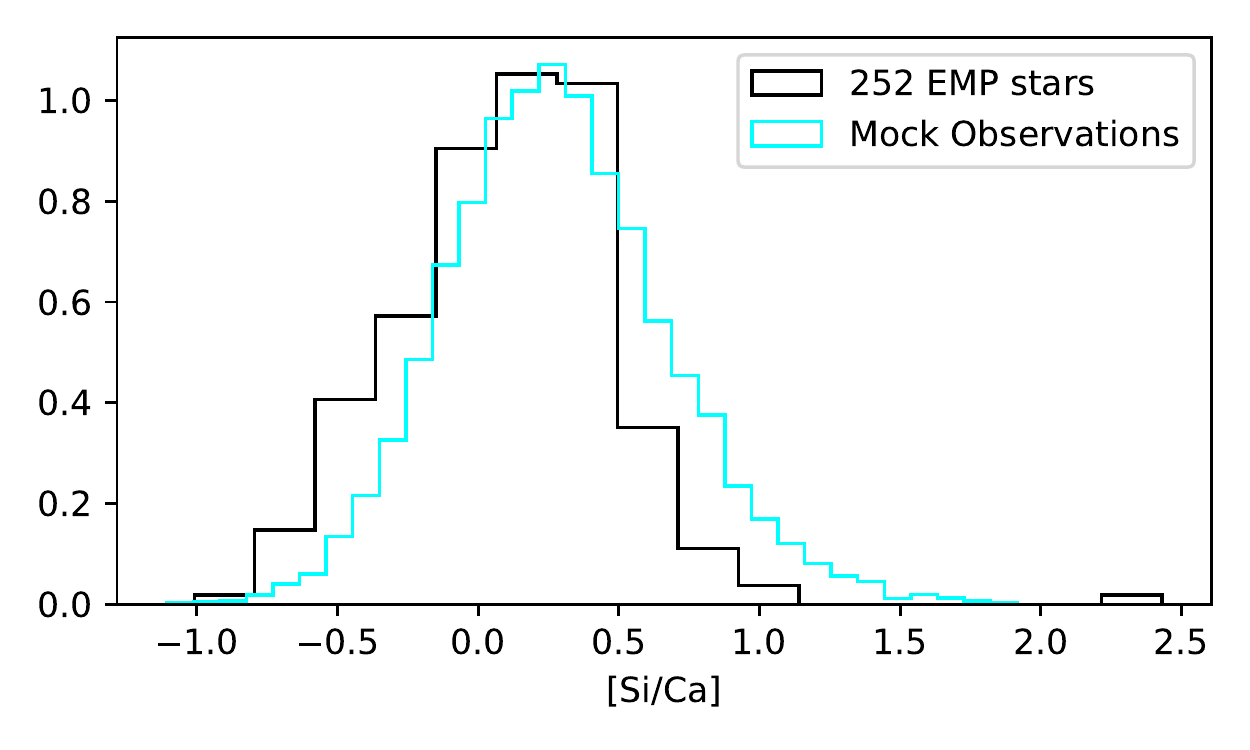}
	\includegraphics[width=0.29\textwidth]{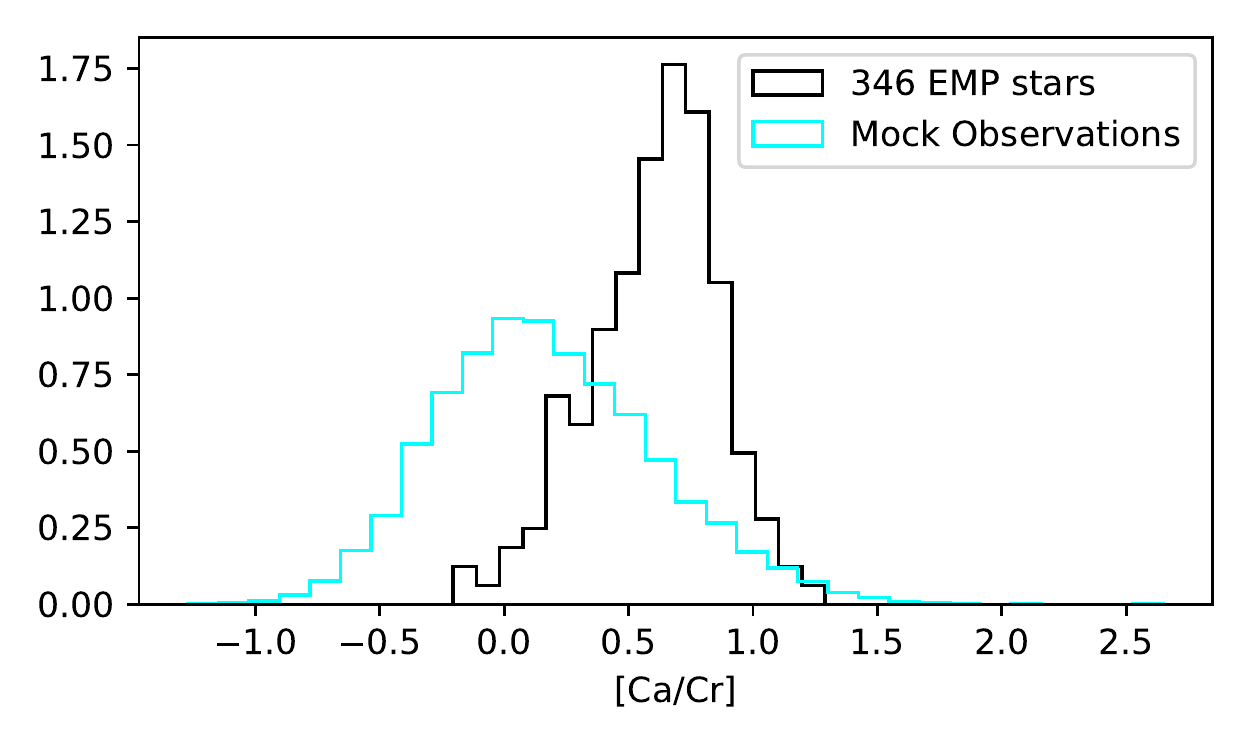}
	\includegraphics[width=0.29\textwidth]{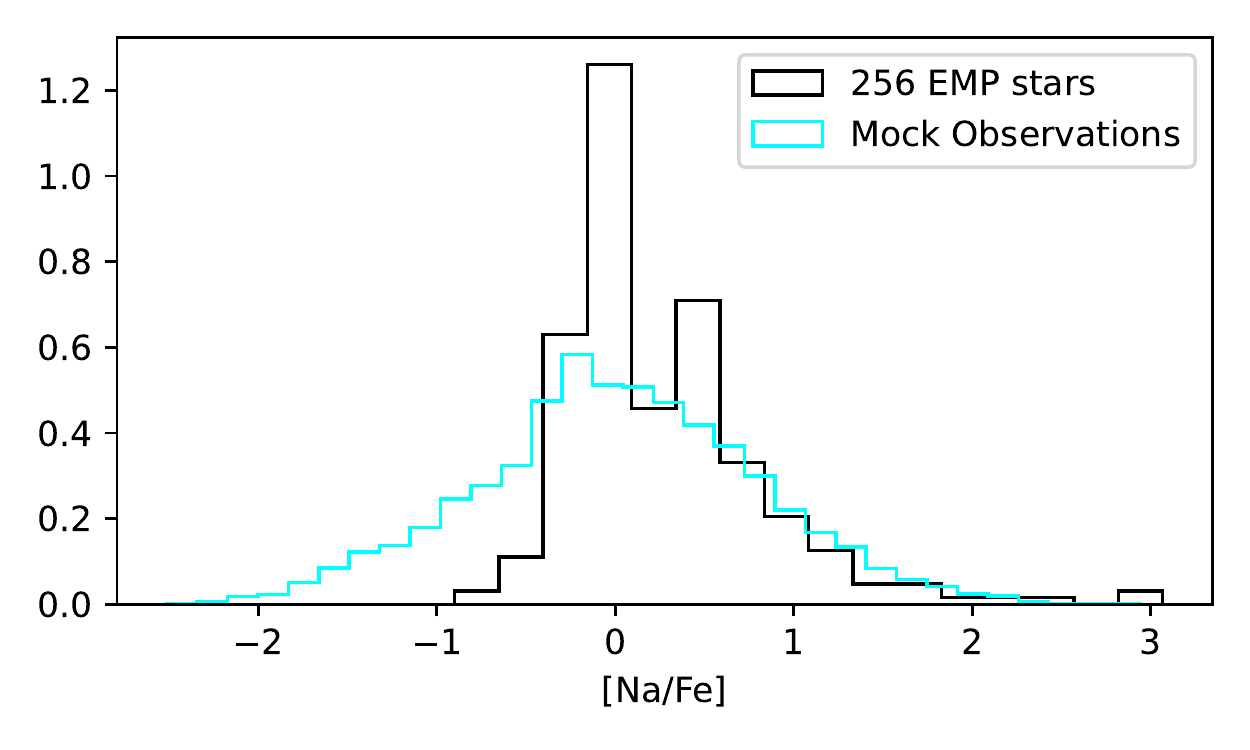}
	\includegraphics[width=0.29\textwidth]{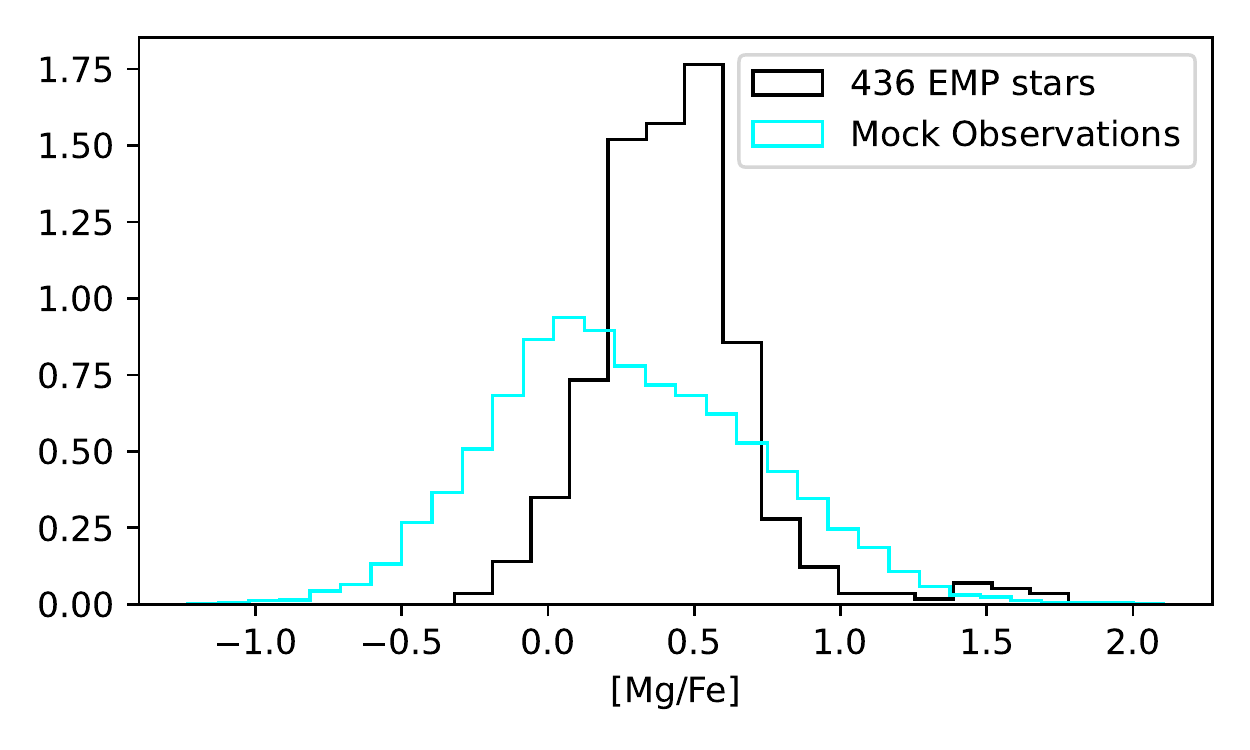}
	\includegraphics[width=0.29\textwidth]{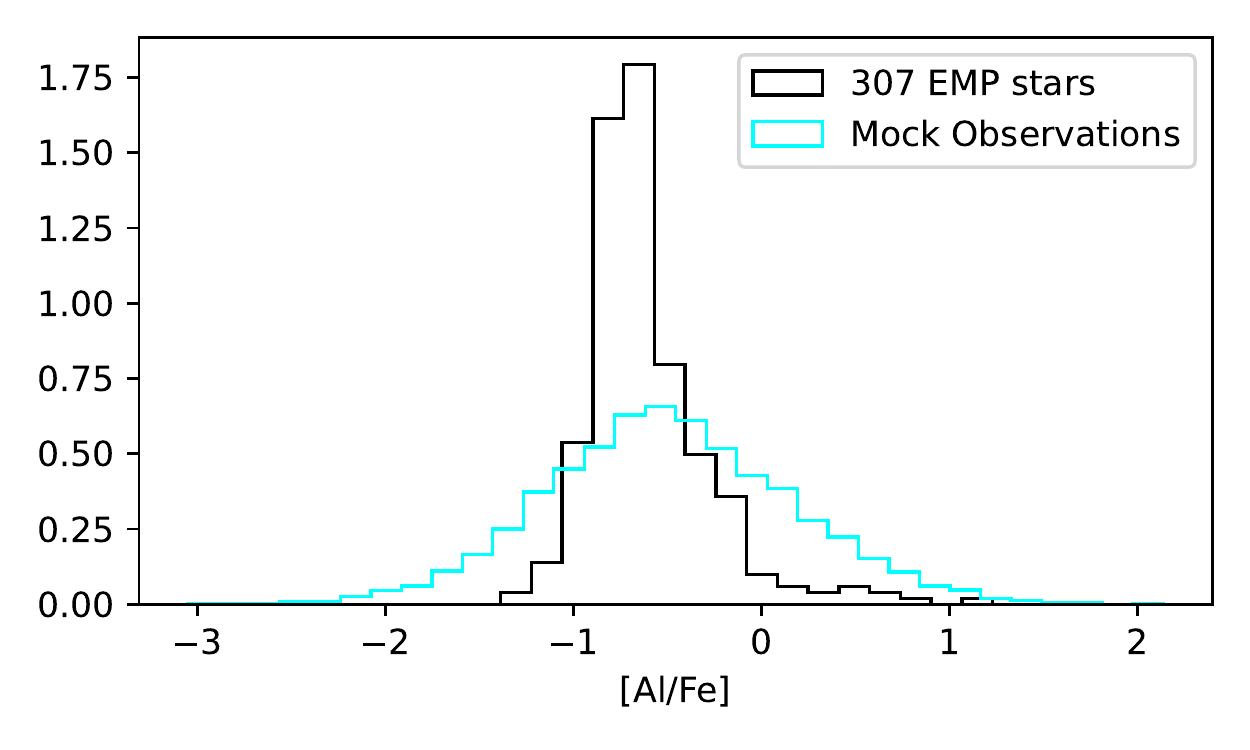}
	\includegraphics[width=0.29\textwidth]{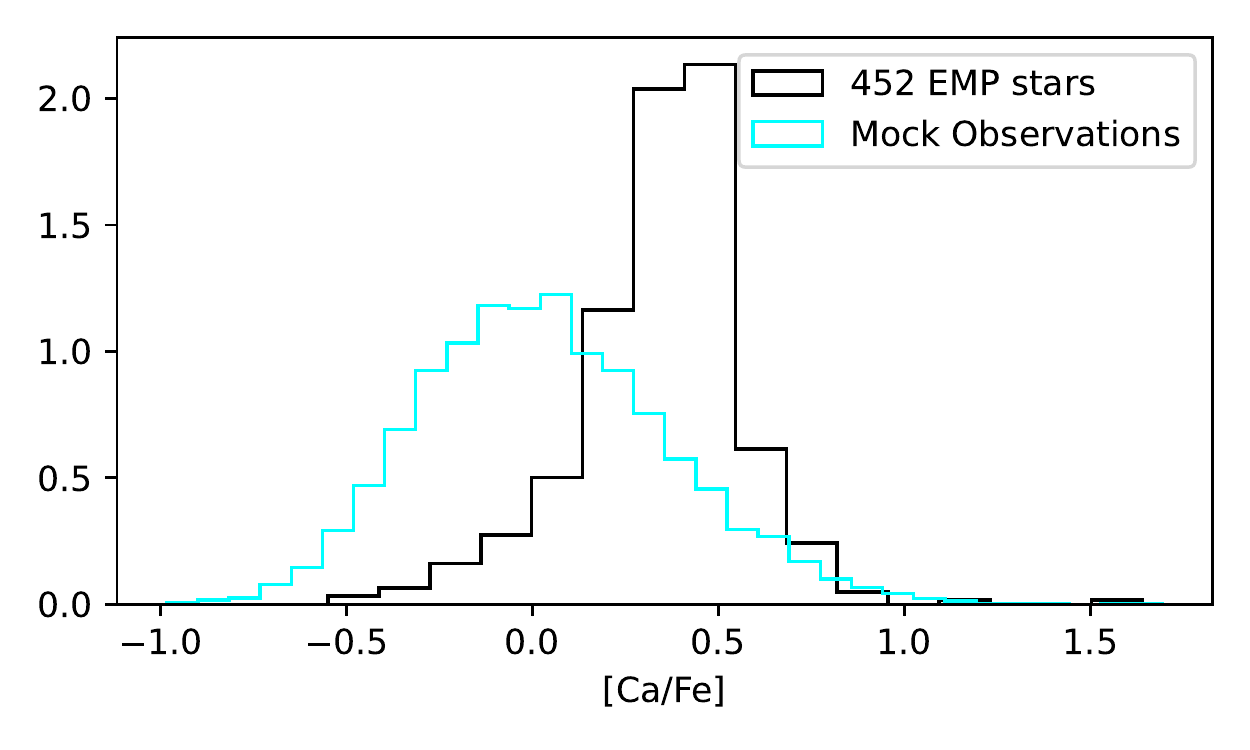}
	\includegraphics[width=0.29\textwidth]{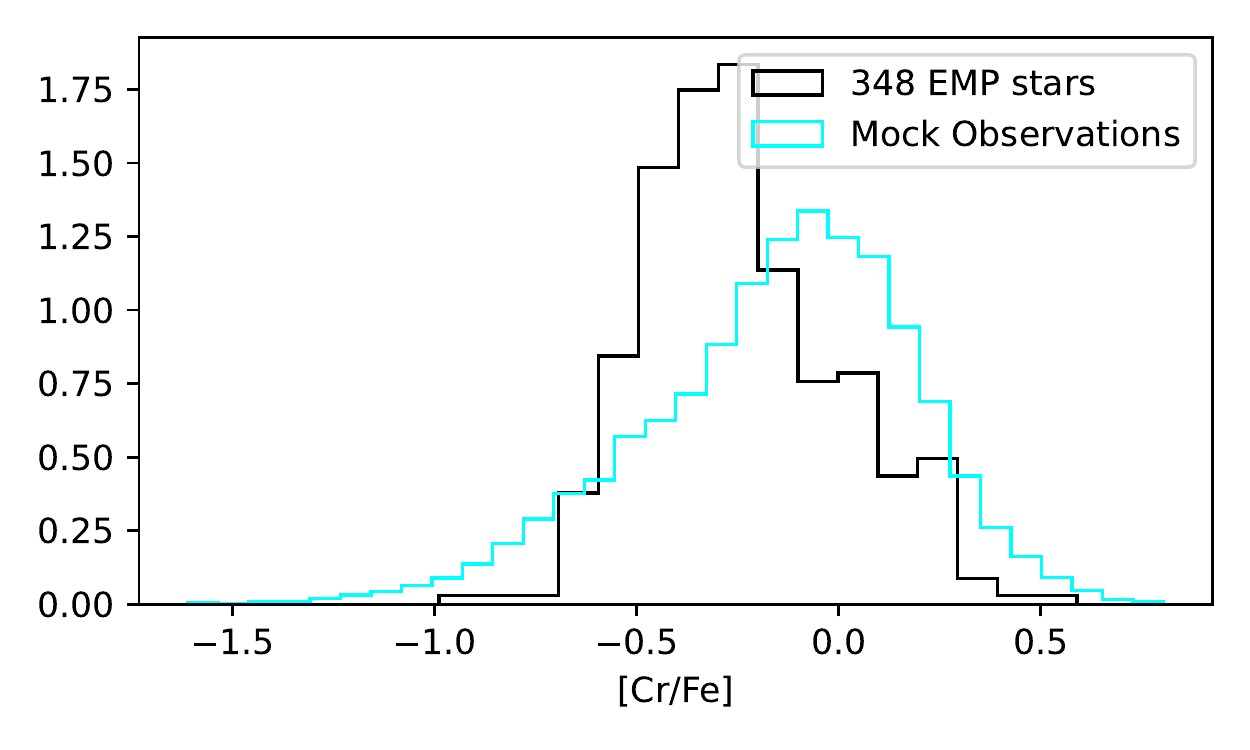}
	\includegraphics[width=0.29\textwidth]{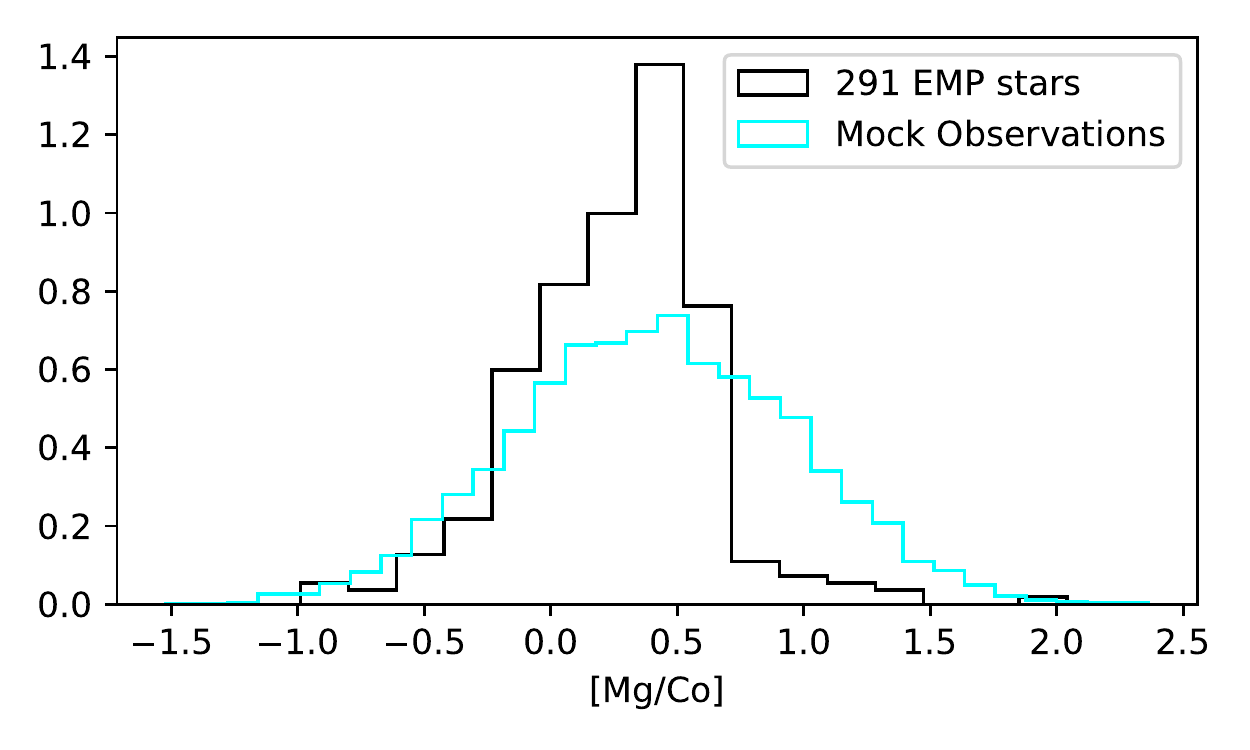}
	\includegraphics[width=0.29\textwidth]{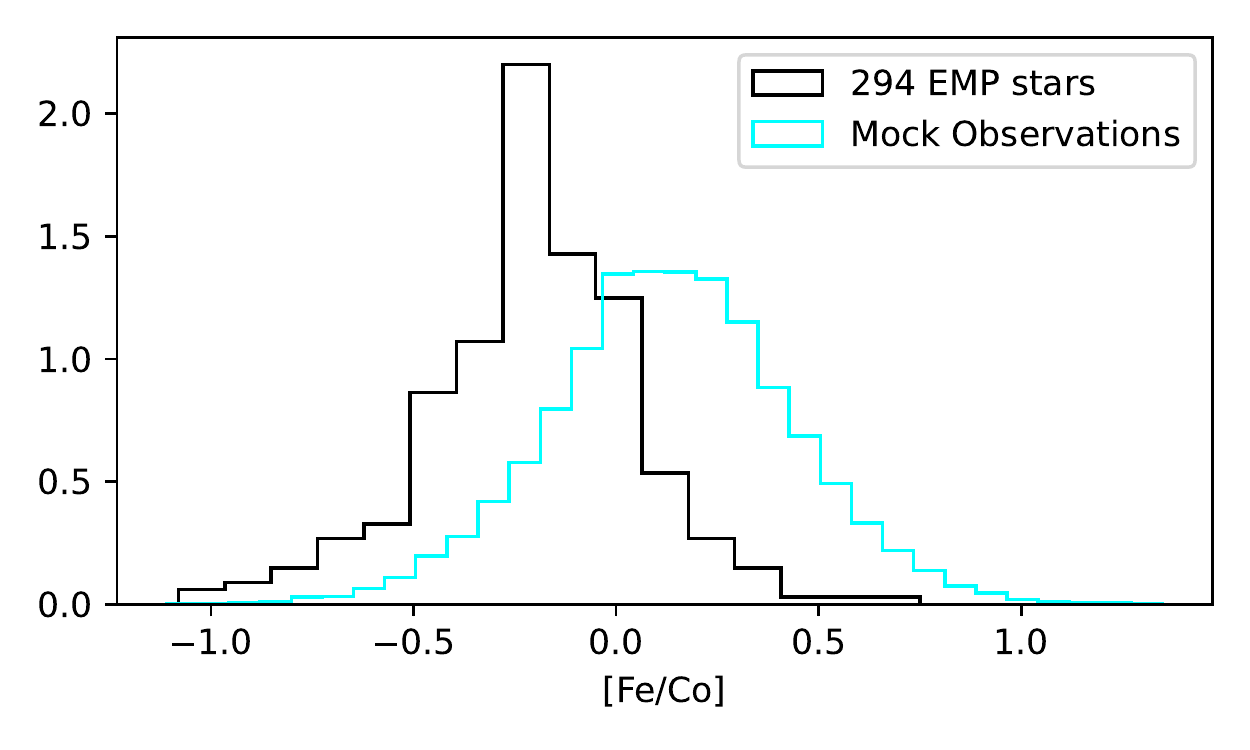}
	\includegraphics[width=0.29\textwidth]{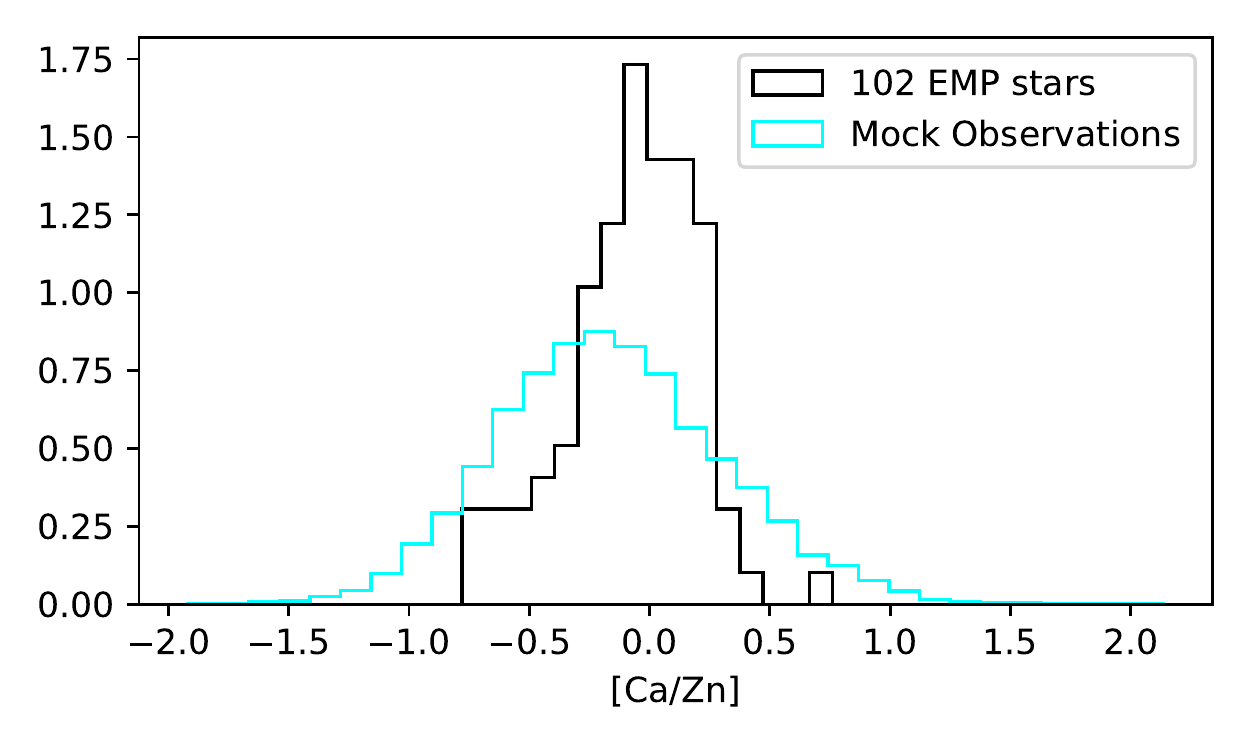}
	\includegraphics[width=0.29\textwidth]{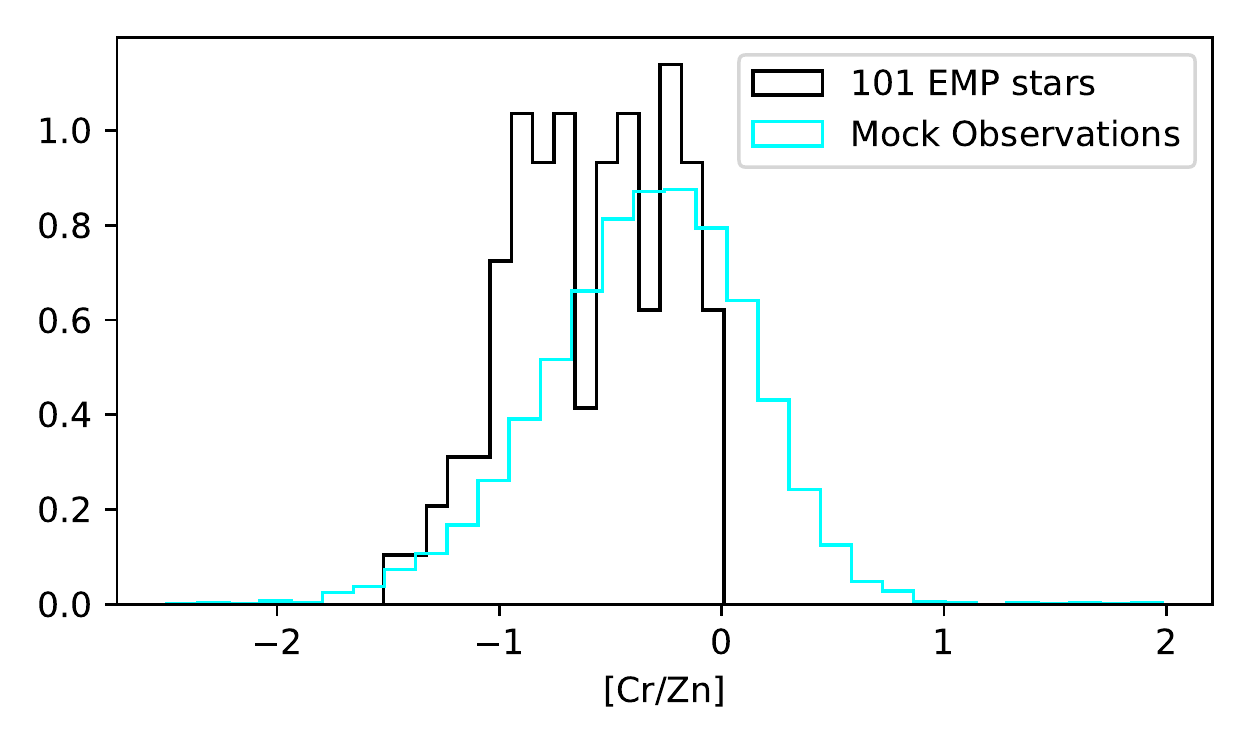}
    \caption{Same as Fig.~\ref{fig:HistTrain}, but for the remaining 21 used abundance ratios.}
    \label{fig:MockHist}
\end{figure}


\bibliography{MyBib}{}
\bibliographystyle{aasjournal}



\end{document}